\documentclass[twocolumn,peerreview,onecolumn,12pt]{IEEEtran}
\usepackage[T1]{fontenc}
\usepackage[latin9]{inputenc}
\usepackage{geometry}
\usepackage{float}
\usepackage{amsmath}
\usepackage{amsthm}
\usepackage{amssymb}
\usepackage{graphicx}
\usepackage[unicode=true,
 bookmarks=true,bookmarksnumbered=true,bookmarksopen=true,bookmarksopenlevel=1,
 breaklinks=false,pdfborder={0 0 0},pdfborderstyle={},backref=false,colorlinks=false]
 {hyperref}
\hypersetup{pdftitle={Your Title},
 pdfauthor={Your Name},
 pdfpagelayout=OneColumn, pdfnewwindow=true, pdfstartview=XYZ, plainpages=false}

\makeatletter

\floatstyle{ruled}
\newfloat{algorithm}{tbp}{loa}
\providecommand{\algorithmname}{Algorithm}
\floatname{algorithm}{\protect\algorithmname}

\theoremstyle{plain}
\newtheorem{thm}{\protect\theoremname}
\theoremstyle{plain}
\newtheorem{lem}[thm]{\protect\lemmaname}

\usepackage[caption=false,font=footnotesize]{subfig}
\usepackage{relsize}
\usepackage{xcolor}
\usepackage{cite}
\usepackage[normalem]{ulem}

\makeatother

\providecommand{\lemmaname}{Lemma}
\providecommand{\theoremname}{Theorem}

\begin{document}

\title{\LARGE{Performance Analysis of a Device-to-Device Offloading Scheme
in a Vehicular Network Environment}\vspace{1cm}
}

\author{\IEEEauthorblockN{Loreto~Pescosolido, Marco~Conti, Andrea Passarella\\
~\\
}\IEEEauthorblockA{\small{Italian National Research Council, Institute for Informatics
and Telematics (CNR-IIT)\\
Via Giuseppe Moruzzi 1, 56124 Pisa, Italy\\
Email: \{loreto.pescosolido, marco.conti, andrea.passarella\}@iit.cnr.it}\vspace{1cm}
}}
\maketitle
\begin{abstract}
We consider a scheme for offloading the delivery of contents to mobile
devices in a vehicular networking scenario. Each content can be delivered
to the requesting device either by a neighboring device or, at the
expiration of a maximum delay, by the network infrastructure nodes.
We compute the analytical expression of the probability that the content
delivery is offloaded through a Device-to-Device (D2D) communication
as a function of the maximum transmission range allowed for D2D communications,
the content popularity, and the vehicles speed. We show that, using
the model, it is possible to identify the optimal maximum transmission
range, which minimizes the total energy consumption (of the infrastructure
plus mobile devices).
\end{abstract}

\begin{IEEEkeywords}
D2D data offloading, vehicular networks, Poisson Point Process\vspace{0.5cm}
\end{IEEEkeywords}

\section{Introduction\setcounter{page}{1}}

In many wireless network scenarios, where mobile devices retrieve
contents by network infrastructure elements (e.g., the eNodeBs in
an LTE or 5G network), Device-to-Device (D2D) communications can be
exploited to obtain relevant network-level performance improvement
in terms of reduction of congestion at the eNodeBs, reduction of the
system-wise transmission energy consumption, and increase of the overall
system spectral efficiency. Most existing system-level studies on
D2D data offloading (see e.g., \cite{Whitbeck2012,Bruno2014,Barbera2014,Rebecchi2016,Rebecchi2015b},
and \cite{Rebecchi2015} for an extensive survey), focus on the congestion
reduction of the cellular network as the main problem to solve through
D2D offloading techniques, and aim to maximize offloading efficiency,
defined as the percentage of contents delivered through D2D, as the
major performance metric. However, in this work, we show that only
maximizing offloading efficiency might result in a very significant
increase of the energy spent by the overall system, i.e., by the eNodeBs
plus the mobile devices, while it is possible to optimize energy efficiency
with a modest reduction of offloading efficiency. 

We show that, by properly setting system parameters at the physical
layer, and in particular the maximum transmission range of the mobile
devices for D2D communications (or, equivalently, the maximum transmit
power for the devices), it is possible to minimize the energy consumption
of the overall system. Intuition suggest that, by increasing the maximum
transmission range of the devices, the probability of offloading keeps
increasing since, at any instant, the number of neighbors of each
node (i.e, the nodes within the transmission range) increases. However,
beyond a certain range, the overall system energy consumption stops
decreasing, since the power required to perform D2D transmissions
starts to become comparable with the power that would be used, to
deliver the contents, by the eNodeBs.

Our goal is to evaluate this effect in quantitative terms, in order
to compute the optimal transmission range which minimizes the overall
system energy consumption, and to understand the impact on offloading
efficiency of operating the system at this optimal operating point.
To achieve this goal, we present an analytical model which captures
application- and scenario-dependent system parameters such as content
popularity, mobility patterns, and vehicle speed, as well as physical
aspects such as the propagation model.

Specifically, we consider a vehicular network scenario and analyze
the performance of the offloading protocol recently proposed in \cite{Bruno2014}.
In this protocol, a node requesting a content can obtain it (i) immediately
through D2D, if a neighbor is caching a copy of that content; (ii)
delayed through D2D, if, by a maximum deadline called content timeout,
it encounters another node caching it; or (iii) delayed through an
eNodeB, if such a node is not encountered within the content timeout.

To the best of our knowledge, in prior system-level studies, the
link with physical parameters, paired with an analytical model which
allows to identify the optimal maximum transmit power for the devices,
has not been considered yet.

The model derived in this work takes into account system model parameters
such as the content popularity and vehicles density and speed, as
well as physical aspects such as the propagation model, and provides
a set of analytical expressions to compute the probability of offloading
a content request, and the average transmit power used to fulfill
it. Both expressions are a function of the maximum D2D transmission
distance, and of the considered system model and channel model. 
Building on this result, we observe that there exist an optimal value
of the maximum D2D transmit power which allows to minimize the overall
system energy consumption. The model is validated through simulations.

The paper is organized as follows. Section~\ref{sec:Related-work}
positions and motivates our work with respect to the existing literature.
Section~\ref{sec:System-model} introduces our system model, the
considered offloading protocol, and a Content Dissemination Management
System (CDMS), operated by the network infrastructure in coordination
with the mobile devices, to execute the offloading protocol. In Section~\ref{sec:results}
we derive the model, and highlight the existence of the optimal operating
point at the system level, in terms of the maximum transmission range
of the devices. In Section~\ref{sec:Performance-evaluation} we validate
the model by means of simulations.  Finally, Section~\ref{sec:Conclusion}
concludes the paper.

\section{Related work and motivation\label{sec:Related-work}}

Techniques for offloading data delivery to D2D-communications have
been extensively investigated in the literature. The interested reader
may want to check, e.g., \cite{Rebecchi2015} for an extensive survey.
From a system-level perspective, the the idea of jointly designing
the offloading and caching/content-injection strategies at the network
layer with information coming from the application layer, such as
content popularity, synchronized or asynchronous requests, and content-sensitive
delay tolerance, has been put forward by several works, under different
assumptions on which contents need to be delivered to which users,
and the absence or presence of delay tolerance. For instance, in \cite{Whitbeck2012,Rebecchi2016},
assuming delay-tolerant applications, and a scenario in which content
delivery mostly relies on D2D-offloading, a strategy for I2D re-injection
of contents in the network is proposed to face temporal content starving
in a certain area. In \cite{Bruno2014}, a CDMS for contents originated
from delay-tolerant applications, suited to vehicular network scenario,
is proposed. In \cite{Barbera2014} the link between social connectivity
and physical connectivity is exploited to select hub nodes in the
network which may assist the cellular infrastructure in the data offloading
process. In \cite{Rebecchi2015b}, in the framework of a content dissemination
problem, the authors propose a mixed I2D-multicast and D2D-relaying
reinforcement-learning-based strategy, which determines which users
should receive the contents through a direct I2D transmission or through
a D2D relaying from a neighboring device. 

At the data-link layer, a class of works is related to organizing
the local D2D topology in order to optimize performance metrics such
as throughput, fairness, and energy/spectral efficiency. For instance,
in \cite{Asadi2017} the authors devise an out-of-band D2D-clustering
strategy, based on coalitional game-theory, aimed at improving these
performance metrics, in a LTE-standard compliant way. In these works,
information related to the application layer, such as the content
popularity, is not considered. Finally, works like \cite{Lin2014,Yang2017}
(amongst many others), aim at devising radio resource allocation strategies,
and/or other physical layer parameters, like coding rates and transmit
power levels, assuming that coexistence among D2D and/or I2D links
are given as an input to the problem. Finally, design and fundamental
limits of D2D caching-based content delivery protocols from an information-theoretic
perspective, are investigated (for an infrastructureless scenario)
in works like \cite{Ji2016a,Ji2016b}, see also the references therein. 

The designs proposed in the works considered above take typically
into account system parameters of the sole referenced architectural
level. For instance, \cite{Whitbeck2012,Bruno2014,Barbera2014,Rebecchi2015b,Rebecchi2016}
only consider system parameters at the application level. Moreover,
in this class of works, an analytical model for the spatiotemporal
stochastic process which controls the positions of the nodes and the
content of their caches, and the related impact on system-level performance,
is lacking, as the performance evaluation is based on simulations
\cite{Bruno2014,Rebecchi2015b,Rebecchi2016,Barbera2014}. On the other
hand, in works in the class of \cite{Asadi2017,Lin2014,Yang2017},
targeting data-link/radio resource management issues, the impact of
content popularity is not considered. Additionally, they typically
assume fully-backlogged traffic and, for the modeling aspects, they
take into account stochastic but static node distribution as in \cite{Lin2014}
or a fixed (and given) one \cite{Yang2017,Asadi2017}. Finally, these
works typically deal more with the coexistence of different D2D links,
rather then dealing with the delivery of individual contents. For
these works, the performance evaluation relies on simulations (\hspace*{-1.2mm}\cite{Lin2014,Yang2017})
or experiments \cite{Asadi2017}. Finally, the results of works in
the class of \cite{Ji2016a,Ji2016b}, are often focusing on scaling
laws and network throughput, assuming a fully backlogged traffic,
but many details of the physical layer are necessarily abstracted
out.

Differently from most of the existing literature, in this work, we
take a full cross-layer approach, including aspects related to the
application layer (such as the content popularity), aspects related
the geographical distribution and mobility of nodes in the network,
and physical layer aspects such as the radio propagation model\footnote{The effect of the radio propagation model is explicitly considered
in our model, but the impact of using different models, due to space
reasons, is not analyzed in this work. A preliminary study of the
impact of different channel models on the performance of the CDMS
considered in this work, has been presented in \cite{BalkanCom2017Offloading}.}. Additionally, we derive an analytical model able to predict the
system performance in terms of a high-level metric such as the offloading
efficiency and metric related to physical parameters such as the system
level average energy consumption. In this way, we can find the optimal
maximum transmit power of the devices, which minimizes the energy
consumption. A cross-layer approach in quite general terms similar
to our one, i.e., which includes the content popularity-related aspects
in the framework of an analytical model based on point process, is
taken in~\cite{Afshang2016}, but in a static scenario and under
completely different assumptions\footnote{In that work, content popularity is related to clusters of users in
the social domain, that are then mapped to physical clusters.}.\vspace{-5mm}

\section{System model and Content Dissemination Management System\label{sec:System-model}}

\subsection{Vehicle arrival and content requests\label{subsec:Vehicle-arrival-and-content-request-model}}

We consider a Region of Interest (ROI) consisting of a street chunk.
Vehicles enter, traverse, and exit the ROI. Each vehicle has onboard
a mobile devices, which can be either a human hand-held device or
part of the vehicle equipment\footnote{The considered offloading protocol is particularly suited to the latter
case since, in that case, the devices extract the energy required
to send cached contents to their neighbors, from a ``virtually''
renewable source like the vehicle battery.}.

We assume that vehicles enter the street from both ends, according
to a homogeneous Temporal Poisson Point Process (TPPP), with vehicle
arrival rate $\lambda_{t}$. The direction from which each new vehicle
enters the street is randomly chosen with equal probability (equal
to $1/2$). Accordingly, the arrival rate of vehicles entering at
one end of the street is $\lambda_{t}/2$. The vehicles traverse the
street at a constant speed. We assume that the speed of each vehicle,
$v$, is a random variable with Probability Density Function (PDF)
$\tilde{p}_{V}(v)$. However, for the derivations in Section~\ref{sec:results},
it is convenient to reformulate the above model by incorporating the
vehicles' motion direction in their speed $v$. This can be done by
including negative speed values. With this formulation, the PDF is
given by
\begin{align}
p_{V}(v)= & \frac{1}{2}u_{(-\infty,0]}(v)\tilde{p}_{V}(-v)+\frac{1}{2}u_{[0,\infty)}(v)\tilde{p}_{V}(v),\label{eq:pdf_speed-1}
\end{align}
where $u_{[x,y]}(\cdot)$ represents the indicator function equal
to one for values of its argument in the interval $[x,y]$, and zero
outside it (the interval is open if one the two extremes is infinite).\\
Most of the results obtained in this work are general with respect
to the speed PDF $p_{V}(v)$. However, for the purpose of performing
simulations to validate the results, it will be useful to consider
a special case. Particularly we shall consider, as a special case,
a uniform distribution of the vehicles speed between two values $v_{a}$
and $v_{b}$, or\linebreak{}
 $\tilde{p}_{V}(v)=\frac{1}{v_{b}-v_{a}}u_{[v_{a},v_{b}]}(v)$. The
two-side PDF (which incorporates the direction) is hence
\begin{equation}
p_{V}(v)=\frac{1}{2(v_{b}-v_{a})}u_{[-v_{b},-v_{a}]}(v)+\frac{1}{2(v_{b}-v_{a})}u_{[v_{a},v_{b}]}(v).\label{eq:PDF_v_special}
\end{equation}

As vehicles enter the ROI, they start requesting contents according
to a given content request process\footnote{The content request process is originated at the application layer.
Here, it is of no importance whether the interest is generated by
a human or by, for instance, an IoT application executed by the software
on a vehicle.}. The content request process is characterized by a content-request
\emph{arrival} process, which defines the time instant at which the
request is generated, and a content-interest probability distribution,
which defines which content is requested. Particularly, we assume
that the devices issue content requests according to a homogeneous
TPPP of constant intensity (rate) $\lambda_{Z}$ content requests
per second, and that contents belong to a finite library $\mathcal{L}$
of size $N_{Z}$. Without loss of generality, we assign an index $z\in\{1,\ldots,N_{Z}\}$
to contents. We assume that content requests follow a given distribution
with Probability Mass Function (PMF) $p_{Z}(z)=Pr\left(Z=z\right)$,
with support $\left[1,\ldots,N_{Z}\right]$. We also assume that successive
requests are independent and identically distributed (iid), and that
requests from different nodes are also iid. Finally, we assume that,
upon obtaining a content, a device keeps it cached for an amount
of time $\tau_{s}$ called \emph{sharing timeout}, so that the cache
occupation is kept limited. 

\subsection{Content Dissemination Management\label{subsec:Content-Dissemination-Management}}

The ROI is served by a set of eNodeBs. During its path within the
ROI, at each instant, each device is associated to an eNodeB, which
is responsible of handling the process of delivering the contents
requested by that device, during the time the devices is in its cell\footnote{The content delivery handling can be also handed over to another eNodeB,
if during the process the requesting device falls within another cell,
see below.}. Similarly to \cite{Whitbeck2012,Bruno2014}, we assume that content
requests have some delay tolerance, i.e., that they must be served
at most within a content timeout $\tau_{c}$. Whenever possible, a
device should obtain a desired content by neighboring or encountered
devices. This is obviously possible if, at the time of content request,
a neighboring device  has the content cached locally. However, as
devices are mobile, there is the chance that, in the event that no
neighbor has the desired content in its cache at the time of request,
the requesting device encounters, later on, another device which does
have the content cached. Only after the content timeout has elapsed,
if the requesting device has not yet obtained the content, it obtains
it from the eNodeB to which it is associated at that time.

\begin{figure}[t]
\centering{}\includegraphics[width=0.7\columnwidth]{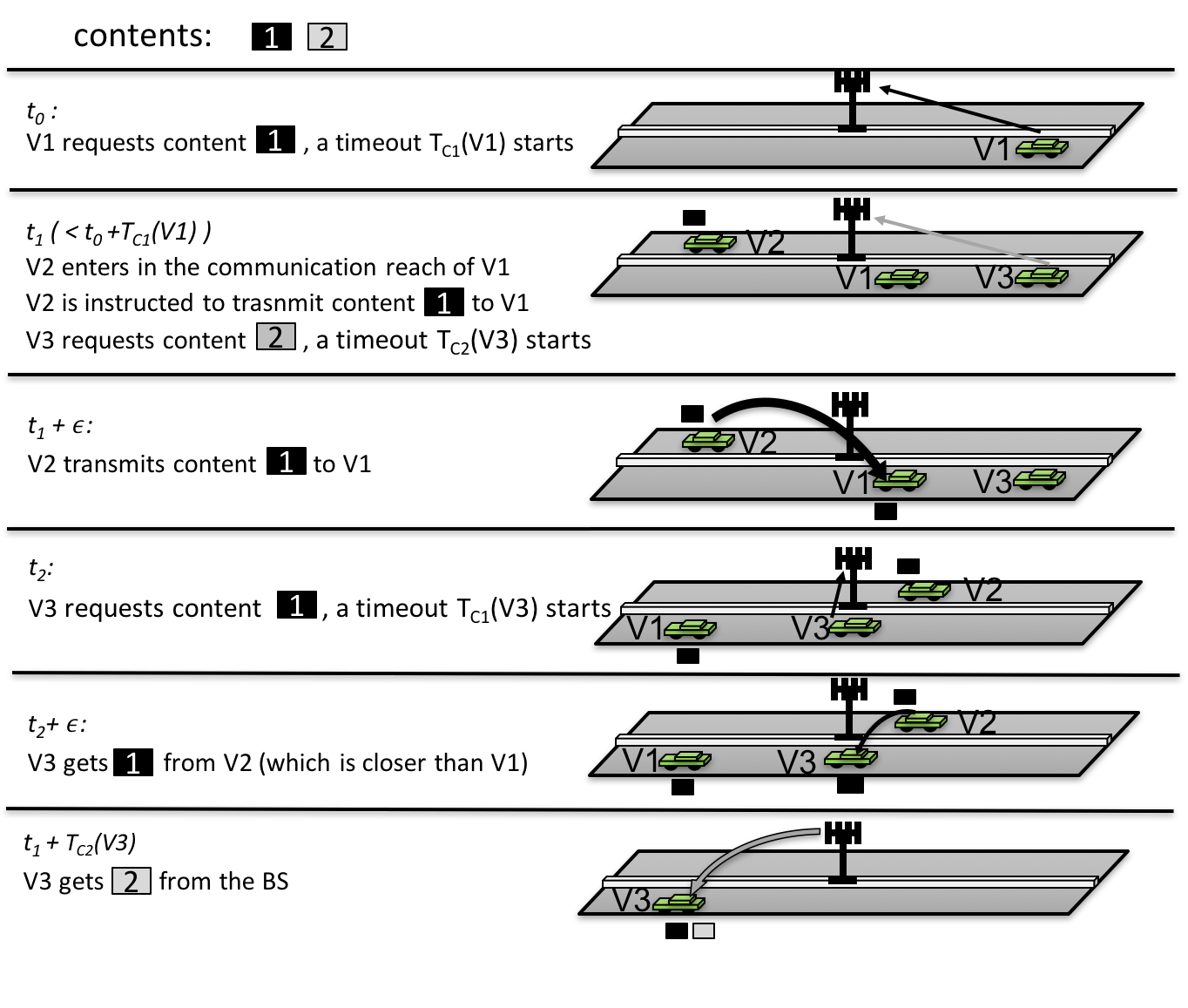}\vspace{-6mm}
\caption{System model and offloading protocol sketch}
\label{fig:sketch}\vspace{-6mm}
\end{figure}

A pictorial representation of this basic idea is provided in Fig.~\ref{fig:sketch},
where the succession of six events at different time instants is represented.
Two vehicles (V1 and V3) request two contents, represented by the
black and grey rectangles. V1 only requests the black content, and
succeeds in obtaining it from the encountered V2 (which has it in
its cache) before the associated content timeout $T_{c1}(V_{1})$
elapses. V3 requests, in successive instants, first the grey content
and then the black content. Similarly to V1, V3 obtains the black
content from V2 before the associated content timeout $T_{c1}(V_{3})$
elapses, while it obtains the grey content from the eNodeB at the
end of the associated content timeout $T_{c2}(V_{3})$ since it has
encountered no device with the content available for D2D-offloading.

The CDMS, essentially, acts on a distributed database (residing at
the eNodeBs) containing the up-to-date list of each node's position,
the list of its neighbors, and the \emph{nominal} channel gain (see
below) between any two neighbors and between each device and the surrounding
eNodeBs. For each device $k$, the list of neighbors, $\mathcal{N}_{k}$,
held at the eNodeB to which device $k$ is associated, is composed
of pairs of the form $(j,r_{j}^{k})$. In this pair, $j$ is the id
of any device which is a neighbor of device $k$, and $r_{j}^{k}$
is a ranking index of device $j$ as ``seen'' by device $k$ on
the basis of a given criterion. In this work, the criterion to establish
if two devices are neighbors, and the ranking of each node's neighbors,
is based on a nominal indicator of the channel quality between the
devices\footnote{In general, the nominal channel quality may be computed, by the CDMS,
on the basis of the positions of the devices, which the eNodeBs are
assumed to know. In this work, we assume that the nominal channel
gain can be computed using any deterministic channel model which relates
the channel gain $g$ to the distance $d$, i.e., a function $g(d)$,
see Subsection~\ref{subsec:Energy-consumption} and Section~\ref{sec:Performance-evaluation}.}. At each eNodeBs, the lists $\mathcal{N}_{k}$ are kept up-to-date
on the basis of Hello messages sent periodically by the devices, containing
a device unique identifier. Each device $k$ has an internal content
cache $\mathcal{C}_{k}$ populated with previously downloaded contents.
At any time, the CDMS also has an index of the contents in each node's
cache, although the CDMS does not necessarily hold a copy of the contents
itself. Time is organized in Control Intervals (CIs). We assume that
the duration of the CIs is much smaller than the content timeout.

Before providing details of the behavior of the CDMS in each CI, it
is worth describing the tasks it performs on a coarse timescale. We
do this through Algorithms~1~and~2, which describe at a high level\footnote{In the pseudocode, the temporal succession of Control Intervals is
not appearing explicitly.} the actions taken on demand, i.e., as a consequence of content requests,
by the devices and the CDMS. We briefly introduce the notation required
for a correct interpretation of the algorithms: $\biguplus\left\{ \mathcal{C}_{j}|\text{condition on }j\right\} $
is used to indicate the union of the caches of devices satisfying
a given condition; $\hat{j}(k,z)$ is used to indicate the device
$j$ that has the best ranking $r_{j}^{k}$ among the neighbors of
device $k$ which have content $z$ in their caches; $j\stackrel{z}{\rightarrow}k$
indicates the transmission of content $z$ from device $j$ to device
$k$. These transmissions are triggered by the CDMS. The remaining
notation used in Algorithms 1-2 is self-explaining.

Upon the generation of a content request, a device (Algorithm~1)
notifies the CDMS that it is interested in that content (step 3),
and then waits for receiving it either from a BS or from a neighbor
(step 4). The system guarantees that the content will be delivered
within the predefined \emph{content timeout}. After the reception
of the content, the device makes it available for other devices that
may request it, for a limited amount of time determined by the \emph{sharing
timeout} (steps 9, and 13-20). 

Algorithm 2 describes the actions taken by the CDMS to handle a content
request. Here, a key point, which effectively allows to increase the
system energy efficiency, is that the CDMS selects the best device
for delivering the content, on the basis of channel quality considerations,
represented by the ranking of each node's neighbors (step 7). If,
however the content cannot be delivered through a D2D communication
within the content timeout, the CDMS uses the eNodeBs to deliver it
(steps 15-20).

The behavior of the CDMS at the CI timescale is as follows. In every
CI, the CDMS schedules which transmissions should be performed, based
on the physical information contained in the above described lists,
and on the ongoing content requests to be handled, including those
whose content timeout has not expired and those whose content timeout
has expired (which need to be fulfilled through I2D transmission).
Specifically, if, at the time of a content request, a requesting device
$k$ has at least one neighboring device with the desired content,
the CDMS selects, out of these neighboring devices, the device $j$
with the best ranking $r_{j}^{k}$, and schedules it to transmit the
content to the requesting device in the following CI. If there are
no neighbors with the requested content available, the first device
encountered by the device $k$, with the desired content cached, is
scheduled by the CDMS to transmit it. Finally, if no device is encountered
within the content timeout, the CDMS schedules the transmission of
the content from the infrastructure. To handle the handover of ongoing
requests originated from a device that crosses a cell border during
the content timeout, adjacent eNodeBs periodically exchange the up
to date status of the ongoing request procedures (see below) of devices
moving across cells\footnote{This information exchange can be performed using high speed fiber
connections, or dedicated radio channels forming a wireless backbone
for the Radio Access Network (RAN).}.
\begin{algorithm}[t]
\caption{{\small{}Actions taken by device $k$ to request content $z$$\protect\phantom{()}$}\texttt{\vspace{1mm}
}}
\label{algo:node_req_handle}
\begin{enumerate}
\item $\mathbf{Upon}$ request for content $z$ from the application layer
\item $\mathbf{Set}$ \texttt{$k$\_content\_received} = $\mathbf{false}$
\item $\mathbf{Send}$ $(k,z)$\texttt{\_cont\_req} to CDMS
\item $\mathbf{while}$ \texttt{$k$\_content\_received ==} $\mathbf{false}$
$\mathbf{do}$\\
~\hspace*{\fill}$\triangleright$ Wait for receiving content $z$,
from a BS or from a neighbor
\item \quad{}$\mathbf{if}$ content $z$ is received $\mathbf{then}$
\item \quad{}\quad{}$\mathbf{Set}$\texttt{ $k$\_content\_received} =
$\mathbf{true}$
\item \quad{}\quad{}$\mathbf{Send}$ $(k,z)\_$ACK to CDMS and/or the
sending device
\item \quad{}\quad{}$\mathbf{Add}$ $z$ to $\mathcal{C}_{k}$
\item \quad{}\quad{}$\mathbf{Set}$ \texttt{$(k,z)$\_sharing\_timeout}
\item \quad{}\quad{}$\mathbf{break}$
\item \quad{}$\mathbf{end\,if}$
\item $\mathbf{end\,while}$
\item $\mathbf{while}$ \texttt{$(k,z)$\_sharing\_timeout} is not expired
$\mathbf{do}$\\
~\hspace*{\fill}$\triangleright$ Available for opportunistic sharing
of content $z$
\item \quad{}$\mathbf{Upon}$ request from CDMS (step 7 of Algorithm \ref{algo_CDMS_req_handle})
\item \quad{}$\mathbf{Send}$ $z$ to device requesting it
\item $\mathbf{end\,while}$
\item $\mathbf{Remove}$ content $z$ from $\mathcal{C}_{k}$
\item $\mathbf{Cancel}$ \texttt{$(k,z)$\_sharing\_timeout}
\end{enumerate}
\end{algorithm}
\begin{algorithm}[t]
\caption{{\small{}Actions taken by CDMS for handling content request $(k,z)$}\texttt{\vspace{1mm}
}}
\label{algo_CDMS_req_handle}
\begin{enumerate}
\item $\mathbf{Upon}$ receiving \texttt{$(k,z)$\_cont\_req}
\item $\mathbf{Set}$ \texttt{$(k,z)$\_served} = $\mathbf{false}$
\item $\mathbf{Set}$ \texttt{$(k,z)$\_content\_timeout} 
\item $\mathbf{while}$ \texttt{$(k,z)$\_content\_timeout} is not expired
$\mathbf{do}$
\item \quad{}$\mathbf{if}$ \texttt{$z\in\biguplus\left\{ \mathcal{C}_{j}|j\in\mathcal{N}_{k}\right\} $}
$\mathbf{then}$
\item \quad{}\quad{}$\mathbf{Identify}$ $\hat{j}(k,z)$
\item \quad{}\quad{}$\mathbf{Trigger}$ transmission $\hat{j}(k,z)\stackrel{z}{\rightarrow}k$
\item \quad{}\quad{}$\mathbf{Wait}$ for \texttt{$(k,z)$\_}ACK
\item \quad{}\quad{}$\mathbf{Upon}$ \texttt{$(k,z)$\_}ACK reception
\item \quad{}\quad{}$\mathbf{Set}$ \texttt{$(k,z)$\_served} = $\mathbf{true}$
\item \quad{}\quad{}$\mathbf{Remove}$ $(k,z)$ from $\mathcal{L}_{\text{req}}$
\item \quad{}\quad{}$\mathbf{break}$
\item \quad{}$\mathbf{end\,if}$
\item $\mathbf{end\,while}$
\item $\mathbf{if}$ \texttt{$(k,z)$\_served} == $\mathbf{false}$
\item \quad{}$\mathbf{Send}$ \texttt{$z$} to \texttt{$k$}
\item \quad{}$\mathbf{Wait}$ for ACK\texttt{\_}$(k,z)$
\item \quad{}$\mathbf{Upon}$ reception of ACK\texttt{\_}$(k,z)$
\item \quad{}$\mathbf{Set}$ \texttt{$(k,z)$\_served} = $\mathbf{true}$
\item $\mathbf{end\,if}$
\item $\mathbf{Cancel}$ \texttt{$(k,z)$\_content\_timeout}
\end{enumerate}
\end{algorithm}
\bigskip{}

\textbf{\emph{Remark:}} The model introduced in Subsection~\ref{subsec:Vehicle-arrival-and-content-request-model}
for the issuing of content requests from a device, does \emph{not}
account for the fact that, at the time of request, the content may
be already present in the cache of the requesting device. If this
is the case, the system is assumed to take the following actions:
(i) the request is labelled as ``repeated'', and therefore the CDMS
does not perform the transmission of the content (either through D2D
or I2D); (ii) the sharing timeout related to that content is reinitialized
to its initial value.

\section{Offloading efficiency and energy consumption\label{sec:results}}

In this section, we compute the probability of offloading through
D2D a non-repeated content request, and the associated transmit power
used on average in each content transmission\footnote{Note that the average energy consumption of the entire system during
a given time interval is simply given by the product of the average
transmit power used for fulfilling a non repeated request, times the
duration of each transmission, times the average number of non-repeated
requests in the interval. \label{fn:power_energy}}. We shall derive expressions of these quantities as a function of
the maximum transmission range allowed to the devices, and of the
content request process and of the vehicles arrival and mobility models
introduced in Subsection~\ref{subsec:Vehicle-arrival-and-content-request-model}.
To derive our analytical results, we first present some preliminary
results obtained by applying standard tools from the theory of temporal
and Spatial Poisson Point Processes (SPPPs), \cite{Daley2008,Moltchanov2012},
(Subsection~\ref{subsec:Preliminary-results})\footnote{It is likely that either of results in Subsection~\ref{subsec:Preliminary-results}
have appeared elsewhere. Nonetheless, for the sake of readability
of the successive derivations, we deem it useful to collect them in
a preliminary subsection.}, then we compute the probability of offloading (Subsection~\ref{subsec:Probability-of-offloading}),
and finally we compute the average transmit power (Subsection~\ref{subsec:Energy-consumption}). 

\subsection{Preliminary results\label{subsec:Preliminary-results}}

We start by proving the following result, which characterizes the
spatial distribution of the vehicles as a function of a given (temporal)
vehicles arrival process.
\begin{lem}
\label{lem:1}Under the assumptions in Subsection~\ref{subsec:Vehicle-arrival-and-content-request-model},
the following results hold true:

1) At any instant, the vehicles are spatially distributed according
to a homogeneous unidimensional\footnote{In our mathematical analysis, we only consider the horizontal coordinate
of the vehicles positions, i.e., we do not take into account that
vehicles moving in opposite directions are located on different lanes
of the street. The comparison of the results of the simulations we
performed to validate our analysis (in which vehicles moving in opposite
direction are placed on different lanes) with the analytical results
shows that the effect of this approximation on the computation of
the offloading efficiency and energy consumption is negligible, see
Section~\ref{sec:Performance-evaluation}.} SPPP with linear density 
\begin{equation}
\rho=\int_{-\infty}^{+\infty}\frac{1}{\left|v\right|}\lambda_{t}p_{V}(v)dv.\label{eq:overall_spatial_density}
\end{equation}
In the special case of uniformly distributed vehicles' speed (\eqref{eq:PDF_v_special}),
\begin{equation}
\rho=\frac{\lambda_{t}\left(\ln v_{b}-\ln v_{a}\right)}{(v_{b}-v_{a})}.\label{eq:overall_spatial_density_special}
\end{equation}

2) Considering a vehicle moving at a specific speed $v^{*}$ on a
straight line, the temporal process of the instants at which the vehicle
encounters\footnote{The ``encountering'' between two vehicles means that they fall within
a range $d_{\max}$ off each other. The instant of the encountering
is the instant at which their distance is exactly equal to $d_{\max}$.} other vehicles, moving at any speed is a homogeneous TPPP with rate
\begin{equation}
\lambda_{e}^{(v^{*})}=\int_{-\infty}^{\infty}\lambda_{t}p_{V}(v)\frac{\left|v^{*}-v\right|}{\left|v\right|}dv.\label{eq:overall_device_encountering_rate_with_general_speed_PDF}
\end{equation}
In the special case \eqref{eq:PDF_v_special},
\begin{equation}
\lambda_{e}^{(v^{*})}=\frac{\lambda_{t}}{(v_{b}-v_{a})}\left(\left|v^{*}\right|\left(\ln\left|v^{*}\right|-\ln v_{a}-1\right)+v_{b}\right).\label{eq:overall_device_encountering_rate_with_general_speed_PDF_SPECIAL}
\end{equation}
\end{lem}
\begin{IEEEproof}
See Appendix A.
\end{IEEEproof}
We focus now on the spatial point process of devices containing a
specific content $z$ in their caches at a given instant, and the
temporal process of the instants at which a point which moves at constant
speed $v^{*}$ encounters vehicles moving at any speed, which have
a specific content $z$ in their caches.
\begin{lem}
\label{lem:2}Under the assumptions in Subsection~\ref{subsec:Vehicle-arrival-and-content-request-model},
the following results hold true:

1) The process of requests for a specific content $z$ issued by a
given device is a homogeneous TPPP, with arrival rate
\begin{equation}
\lambda_{z}=p_{Z}(z)\lambda_{Z}.\label{eq:arrival-rate-of-TPPP-req_for_content_z-1}
\end{equation}

2) At any instant, the probability $Pr\left(\mathcal{C}\ni z\right)$
that the cache of a generic device contains a specific content $z$
is upper and lower bounded as follows
\begin{align}
1-e^{-\lambda_{z}(\tau_{s}-\tau_{c})}\leq & \,Pr\left(\mathcal{C}\ni z\right)\leq1-e^{-\lambda_{z}\tau_{s}}.\label{eq:Prob_content_z_in_cache-1}
\end{align}

3) At a given instant, the spatial process of the position of the
devices containing a specific content $z$ in their caches can be
very well approximated by a homogeneous SPPP, with linear density
$\rho_{z}$ tightly lower bounded as in
\begin{equation}
\rho_{z}\gtrsim\rho\left(1-e^{-\lambda_{z}(\tau_{s}-\tau_{c})}\right),\label{eq:SPPP-geo-content-distribution-2}
\end{equation}
where $\rho$ is given by~\eqref{eq:overall_spatial_density} or
\eqref{eq:overall_spatial_density_special}.

4) Consider a vehicle moving at speed $v^{*}$. The temporal process
of devices, that have a specific content $z$ in their caches, encountered
by the vehicle, can be very well approximated by a a homogeneous TPPP
with encountering rate tightly lower bounded as
\begin{equation}
\lambda_{e}^{(v^{*},z)}\gtrsim\lambda_{e}^{(v^{*})}\left(1-e^{-\lambda_{z}(\tau_{s}-\tau_{c})}\right),\label{eq:encountering_rate_with_devices_with_content_z-1}
\end{equation}
where $\lambda_{e}^{(v^{*})}$ is given by \eqref{eq:overall_device_encountering_rate_with_general_speed_PDF},
or \eqref{eq:overall_device_encountering_rate_with_general_speed_PDF_SPECIAL}
in the special case \eqref{eq:PDF_v_special}.
\end{lem}
\begin{IEEEproof}
See Appendix A.
\end{IEEEproof}
Note that, since in practical scenarios we may reasonably assume that
$\tau_{s}\gg\tau_{c}$, the bounds in \eqref{eq:Prob_content_z_in_cache-1}
are very tight. As a result, in practice, the lower bounds~\eqref{eq:SPPP-geo-content-distribution-2}~and~\eqref{eq:encountering_rate_with_devices_with_content_z-1}
can be considered as very accurate approximations\footnote{Both expressions \eqref{eq:SPPP-geo-content-distribution-2}~and~\eqref{eq:encountering_rate_with_devices_with_content_z-1}
result from approximating $Pr\left(\mathcal{C}\ni z\right)$ with
its lower bound in \eqref{eq:Prob_content_z_in_cache-1}. Using the
upper bound in \eqref{eq:Prob_content_z_in_cache-1} would still entail
an accurate approximation (consisting in tight upper bounds instead
of lower bounds) of the density and rate appearing in \eqref{eq:SPPP-geo-content-distribution-2}~and~\eqref{eq:encountering_rate_with_devices_with_content_z-1},
respectively. For practical purposes, the impact of using either of
the two bounds is negligible. Selecting the lower bound represents
a conservative choice for the performance evaluation, since it tends
to underestimate (in a negligible way) the probability of offloading
the content requests.\label{fn:Both-expressions-and}}. In the following, we will use the notation ``$\backsimeq$''
in all the Equations that stem from \eqref{eq:SPPP-geo-content-distribution-2}~and~\eqref{eq:encountering_rate_with_devices_with_content_z-1},
as a convention to state that the approximation is quite accurate
since it is supported by tight upper and lower bounds.

\subsection{Probability of content delivery offloading\label{subsec:Probability-of-offloading}}

The results obtained in Subsection~\ref{subsec:Preliminary-results}
allow us to compute the probability that the fulfilling of a content
request is offloaded to a D2D transmission among nearby devices. We
shall compute this probability as a function of the maximum nominal
transmission range of the devices, indicated in the following with
$d_{\max}$. If two devices, at a given instant, are closer than $d_{\max}$,
they are considered to be neighbors. Note that $d_{\max}$ is tightly
related to physical layer parameters, such as transmit power and information
rate, which play a major role in the determination of the system energy
consumption, see Subsection~\ref{subsec:Energy-consumption}.

In the following, in using the terminology ``probability of offloading'',
\emph{we always refer to the probability conditioned on the fact that
the request is not repeated} (see the final remark in Subsection~\ref{subsec:Content-Dissemination-Management}).
To avoid using an excessively cumbersome notation, we omit this conditioning
from the notation of this subsection up to Eq.~\eqref{eq:Prob_NON_offloading}.

To compute the probability of offloading (of a non-repeated request),
we start computing the probability of offloading a non-repeated request,
further conditioned on the fact that the requested content is a specific
one, say $z$. We first compute the probability that the request is
fulfilled \emph{immediately. }We indicate the probability of this
event with $Pr^{(d_{\max})}\left(\text{off.imm}\mid z\right)$. Now,
the request can be fulfilled \emph{immediately,} through a D2D transmission,
if at least one neighbor of the requesting node has content $z$ in
its cache at the time of request. This is equivalent to say that the
\emph{closest neighbor}\footnote{Under the assumption that the neighbors ranking is performed on the
basis of a distance-based criterion.} has content $z$ in its cache at the time of request. This event
is determined by the SPPP of the devices containing content $z$,
which is a homogeneous SPPP with intensity $\rho_{z}$ given by \eqref{eq:SPPP-geo-content-distribution-2},
and by the maximum nominal transmission range $d_{\max}$. In fact,
$Pr^{(d_{\max})}\left(\text{off.imm}\mid z\right)$ coincides with
the probability that the closest point of the homogeneous SPPP of
the devices containing content $z$, is at a distance less than $d_{\max}$
at the time of request. It is well known that the Cumulative Distribution
Function (CDF) of the ``closest neighbor distance'' $d_{\text{cn}}$
determined by a unidimensional homogeneous SPPP with (linear) density
$\tilde{\rho}$, is given by $F_{\text{cn}}(d)\triangleq Pr\left(d_{\text{cn}}\leq d\right)=1-e^{-\tilde{\rho}2d}$,
\cite{Daley2008,Moltchanov2012}. Accordingly, we obtain\vspace{-1mm}
\begin{equation}
Pr^{(d_{\max})}\left(\text{off.imm}\mid z\right)=1-e^{-2d_{\max}\rho_{z}}.\label{eq:Prob_offloading_immediate_param_z}
\end{equation}

Using \eqref{eq:arrival-rate-of-TPPP-req_for_content_z-1} in the
expressions of the upper and lower bounds in \eqref{eq:Prob_content_z_in_cache-1},
plugging the lower bound in \eqref{eq:Prob_content_z_in_cache-1}
to compute $\rho_{z}$ from \eqref{eq:SPPP-geo-content-distribution-2},
and using \eqref{eq:SPPP-geo-content-distribution-2} in \eqref{eq:Prob_offloading_immediate_param_z},
we obtain the following expression for the probability of immediate
offloading of a non-repeated request of a specific content $z$:\vspace{-1mm}
\begin{align}
Pr^{(d_{\max})}\left(\text{off.imm}\mid z\right)\backsimeq & 1-e^{-2d_{\max}\cdot\left(1-e^{-p_{Z}(z)\lambda_{Z}\cdot(\tau_{s}-\tau_{c})}\right)\rho},\label{eq:prob_offloading_content_z_lower_bound}
\end{align}
where, in the special case of $p_{V}(v)$ given by \eqref{eq:PDF_v_special},
$\rho$ can be replaced by $\lambda_{t}\left(\ln v_{b}-\ln v_{a}\right)/(v_{b}-v_{a})$,
see Eq.~\eqref{eq:overall_device_encountering_rate_with_general_speed_PDF_SPECIAL}.

Next, we compute the probability that the request is still fulfilled
through a D2D transmission, but the content is obtained by a device
which, during the content timeout following the request, comes within
a range $d_{\max}$ off the requesting device, i.e., it is encountered
by it. We first compute such probability for a requesting device moving
at a specific speed $v$.

First, consider a vehicle moving at speed $v$ and an instant $t_{0}$.
The probability that the first device with a given content $z$ in
its cache comes within a range $d_{\max}$ off the requesting device,
starting from $t_{0}$, within an interval of duration equal to the
content timeout $\tau_{c}$, is given by\vspace{-1mm}
\begin{align}
Pr\left(\text{enc}\mid z,v\right) & =1-e^{-\lambda_{e}^{(v,z)}\tau_{c}}\backsimeq1-e^{-\lambda_{e}^{(v)}\left(1-e^{-\lambda_{z}(\tau_{s}-\tau_{c})}\right)\tau_{c}},\label{eq:Prob_ecounter_param_z_param_v}
\end{align}
where $\lambda_{e}^{(v,z)}$ has the expression \eqref{eq:encountering_rate_with_devices_with_content_z-1}.
In the special case \eqref{eq:PDF_v_special}, we have
\begin{align}
Pr\left(\vphantom{\text{enc}\mid z,v}\right. & \hspace{-1.5mm}\left.\text{enc}\mid z,v\right)\backsimeq1-e^{-\frac{\lambda_{t}}{(v_{b}-v_{a})}\left(\left|v\right|\left(\ln\left|v\right|-\ln v_{a}-1\right)+v_{b}\right)\left(1-e^{-\lambda_{z}(\tau_{s}-\tau_{c})}\right)\tau_{c}}.\label{eq:Prob_encounter_param_z_param_v_SPECIAL}
\end{align}
We observe that this probability does not depend on $d_{\max}$.

Now, the probability that a non-repeated request for a specific content
$z$, issued by a vehicle moving at speed $v$, is fulfilled through
a D2D transmission by a device encountered during the content timeout
following the request, is given by the probability that the request
has \emph{not} been fulfilled immediately, $\left(1-P^{(d_{\max})}\left(\text{off.imm}\mid z\right)\right)$,
times the probability \eqref{eq:Prob_ecounter_param_z_param_v} of
encountering, within the content timeout following the request, a
device with the desired content $z$ in its cache, i.e.
\begin{equation}
\small{Pr^{(d_{\max})}\left(\text{off.del}\mid z,v\right)\hspace{-0.5mm}=\hspace{-0.5mm}\left(1\hspace{-1mm}-\hspace{-1mm}Pr^{(d_{\max})}\left(\text{off.imm}\mid z\right)\right)\hspace{-1mm}Pr\left(\text{enc}\mid z,v\right).}\label{eq:Prob_off_delayed_(z,v)}
\end{equation}

This quantity depends on $d_{\max}$ only through $Pr^{(d_{\max})}\left(\text{off.imm}\mid z\right)$.

Removing the dependence on the speed $v$ at which the requesting
device is moving, we can compute the probability, indicated with $Pr^{(d_{\max})}\left(\text{off.del}\mid z\right)$,
that a non-repeated request for a specific content $z$, issued by
a vehicle moving at \emph{any} speed, is fulfilled through a D2D transmission
by a device encountered during the content timeout following the request.
By the law of total probability, we get
\begin{align}
Pr^{(d_{\max})}\left(\text{off.del}\mid z\right)= & \int_{-\infty}^{\infty}Pr^{(d_{\max})}\left(\text{off.del}\mid z,v\right)p_{V}\left(v\right)dv\nonumber \\
\small{}= & \left(1-Pr^{(d_{\max})}\left(\text{off.}\right.\right.\hspace{4mm}\hspace{-4mm}\left.\left.\hspace{-1.5mm}\text{imm}\hspace{-1mm}\mid\hspace{-1mm}z\right)\vphantom{1-Pr^{(d_{\max})}\left(\text{off.}\right.}\right)\hspace{-1.5mm}\int_{-\infty}^{\infty}\hspace{-5mm}Pr\left(\text{enc}\mid z,v\right)p_{V}\left(v\right)dv.\label{eq:Prob_offloading_delayed_param_z}
\end{align}
This expression is general with respect to the speed PDF $p_{V}(v)$.
In a practical scenario, to quantitatively evaluate the probability
of offloading, a specific model for $p_{V}(v)$ should be provided
in input to \eqref{eq:Prob_offloading_delayed_param_z}. For instance,
taking $p_{V}(v)$ as in \eqref{eq:PDF_v_special}, it is straightforward
to show that \eqref{eq:Prob_offloading_delayed_param_z} becomes
\begin{align}
Pr^{(d_{\max})}\left(\text{off.del}\mid z\right)= & \left(1-Pr^{(d_{\max})}\left(\text{off.imm}\mid z\right)\right)\label{eq:Prob_offloading_delayed_param_z_SPECIAL}\\
 & \cdot\left(1-\frac{e^{-\frac{\lambda_{t}v_{b}Pr\left(z\in\mathcal{C}\right)\tau_{c}}{(v_{b}-v_{a})}}}{(v_{b}-v_{a})}\right.\left.\int_{v_{a}}^{v_{b}}e^{-\frac{\lambda_{t}Pr\left(z\in\mathcal{C}\right)\tau_{c}}{(v_{b}-v_{a})}v\left(\ln v-\ln v_{a}-1\right)}dv\vphantom{\frac{e^{-\frac{\lambda_{t}v_{b}Pr\left(z\in\mathcal{C}\right)\tau_{c}}{(v_{b}-v_{a})}}}{(v_{b}-v_{a})}}\right).\nonumber 
\end{align}

The total probability that a non-repeated request for content $z$
is fulfilled through offloading is obviously given by
\[
\small{Pr^{(d_{\max})}\left(\text{off}\hspace{-1mm}\mid\hspace{-1mm}z\right)=Pr^{(d_{\max})}\left(\text{off.imm}\hspace{-1mm}\mid\hspace{-1mm}z\right)+Pr^{(d_{\max})}\left(\text{off.del}\hspace{-1mm}\mid\hspace{-1mm}z\right)}
\]

Finally, the probability that a non-repeated request for content $z$
is fulfilled through an I2D transmission, i.e., it is not offloaded,
is given by
\begin{align}
Pr^{(d_{\text{max}})}\left(\text{non-off}\mid z\right) & =1\hspace{-1mm}-\hspace{-1mm}Pr^{(d_{\max})}\left(\text{off.imm}\mid z\right)-\hspace{-1mm}Pr^{(d_{\max})}\left(\text{off.del}\mid z\right).\label{eq:Prob_NON_offloading}
\end{align}
We now proceed removing the dependence on the requested content~$z$.
First, we prove the following
\begin{lem}
\label{lem:3}Under the assumptions in Subsection~\ref{subsec:Vehicle-arrival-and-content-request-model},
the probability that a content request is not repeated, is given by
\begin{equation}
Pr\left(\mathrm{NR}\right)=\sum_{z}Pr\left(Z=z\right)Pr\left(\mathcal{C}\not\ni z\right),\label{eq:Prob_non_repeated_req}
\end{equation}
and the probability that the content $Z$ requested in a content request,
conditioned to the fact that the request is not repeated, is given
by
\begin{equation}
p_{Z}\left(z\mid\mathrm{NR}\right)=\frac{Pr\left(Z=z\right)Pr\left(\mathcal{C}\not\ni z\right)}{\sum_{z\in\mathcal{L}}Pr\left(Z=z\right)Pr\left(\mathcal{C}\not\ni z\right)}.\label{eq:Prob_Z_conditioned_NR}
\end{equation}
\end{lem}
\begin{IEEEproof}
See the Appendix A.
\end{IEEEproof}
Finally, we obtain the following
\begin{thm}
The probability of offloading for a non-repeated request, irrespective
of the requested content, is given by
\begin{align}
Pr^{(d_{\max})}\left(\mathrm{off}\mid\mathrm{NR}\right) & =\sum_{z\in\text{\ensuremath{\mathcal{L}}}}p_{Z}\left(z\mid\mathrm{NR}\right)Pr^{(d_{\max})}\left(\mathrm{off}\mid z\right)\label{eq:prob_OFF}\\
=\sum_{z\in\text{\ensuremath{\mathcal{L}}}} & \frac{p_{Z}\left(z\right)Pr\left(\mathcal{C}\not\ni z\right)}{\sum_{z\in\mathcal{L}}p_{Z}\left(z\right)Pr\left(\mathcal{C}\not\ni z\right)}Pr^{(d_{\max})}\left(\mathrm{off}\mid z\right).\nonumber 
\end{align}
\end{thm}
\begin{IEEEproof}
This comes straightforward from applying the law of total probability,
and replacing the probability of requesting $z$ (conditioned to the
event that the request is not repeated) with Eq.~\eqref{eq:Prob_Z_conditioned_NR}.
Eq.~\eqref{eq:prob_OFF} expresses the law of total probability applied
to the event of offloading a non-repeated request. i.e., it simply
states that the probability is the sum, over all the possible realizations
$z$ of the requested content $Z$, of the probability of the offloading
event conditioned to each specific realization $z$, or $Pr^{(d_{\max})}\left(\text{off}\mid z\right)$,
weighted by the probability that the requested content is $z$ (conditioned
to the fact that the request is not repeated), or $p_{Z}\left(z\mid\text{NR}\right)$.
The specific expression of $p_{Z}\left(z\mid\text{NR}\right)$ is
given by~\eqref{eq:Prob_Z_conditioned_NR} in Lemma~\ref{lem:3}.
\end{IEEEproof}

\subsection{Energy consumption minimization\label{subsec:Energy-consumption}}

We consider, without loss of generality an LTE-like multi-carrier
communication system. A set of Physical Resource Blocks (PRBs), corresponding
to the elements of a time-frequency grid, is allocated to each communication,
on the basis of the size of the content that needs to be transmitted,
see below. In the following, we assume a fixed content size of $D$
bits. As stated in Subsection~\ref{subsec:Content-Dissemination-Management}
the CDMS is aware of the nominal (scalar) channel gain between any
D2D or I2D pair. Therefore, power is allocated uniformly over the
subcarriers. Let $\overline{e}$ be a nominal target normalized (i.e.,
measured in bps/Hz) information rate that a link is required to be
able support (in this work, this is consider a fixed system parameter).
Let $\mathcal{P}^{(w_{c})}$ be the transmit power allocated \emph{on
each subcarrier}, $g(d)$ be a generic monotonically decreasing propagation
loss formula which relates distance $d$ to the nominal channel gain
$g$ in a deterministic way, $w_{c}$ the subcarrier spacing, $F_{rc}$
the noise figure at the receiver, $N_{0}$ the thermal noise power
spectral density, and $\sigma_{c}^{2}=w_{c}F_{rc}N_{0}$ the noise
power on each subcarrier. Let $\mathcal{P}_{tx}^{(w_{c})}$ be the
transmit power allocated on each subcarrier, and let the nominal channel
gain be $g(d)$. We define the normalized nominal information rate
(measured in bps/Hz) $e$ as the Shannon capacity on that subcarrier
divided by the subcarrier width, or
\begin{equation}
\small{e=\frac{1}{w_{c}}w_{c}\log_{2}\left(1+\frac{\mathcal{P}_{tx}^{(w_{c})}g(d)}{\sigma_{c}^{2}}\right)=\log_{2}\left(1+\frac{\mathcal{P}_{tx}^{(w_{c})}g(d)}{\sigma_{c}^{2}}\right).}\label{eq:normalized-nominal-information-rate-per subcarrier-1}
\end{equation}
We assume that, to transmit to a receiver located $d$ meters away,
the transmitter sets the the transmit power over each subcarrier to
\begin{equation}
\mathcal{P}_{tx,w_{c}}^{(\bar{e})}(d)=\frac{1}{g(d)}\sigma_{c}^{2}\left(2^{\bar{e}}-1\right).\label{eq:tx_power_as_a_function_of_distance}
\end{equation}
This is obtained by inverting \eqref{eq:normalized-nominal-information-rate-per subcarrier-1}
with respect to $\mathcal{P}_{tx,w_{c}}^{(\bar{e})}$, with the objective
to match the target nominal normalized information rate $\overline{e}$.

If a non-repeated request for content $z$ if fulfilled immediately
through offloading, the distance at which the transmitter is located
is the distance of the nearest neighbor (with content $z$ in its
cache), \emph{conditioned to the fact that the nearest neighbor is
within a range $d_{\max}$ off the requesting device. }Let us indicate
the nearest neighbor distance with the random variable $D$. The
required transmit power to fulfill a non-repeated request is the random
variable resulting from the transformation of the random variable
$D$ to the random variable $Y_{\text{NR,off,im}}$ defined as
\begin{equation}
\small{Y_{\text{NR,off,im}}\triangleq\mathcal{P}_{tx}^{(\bar{e})}\left(D\right)=\frac{1}{g\left(D\right)}\sigma_{c}^{2}\left(2^{\bar{e}}-1\right).}\label{eq:Y_NR_OFF_IM_DEFINITION}
\end{equation}
With relatively straightforward integral calculus steps, it can be
showed that the CDF $F_{Y,\text{NR,off,im}}\left(y;z\right)$, the PDF
$p_{Y,\text{NR,off,im}}(y;z)$, and average value $\overline{Y}_{\text{NR,off,im}}^{(z)}\left(d_{\max}\right)$
of $Y_{\text{NR,off,im}}$, computed as a function of the system parameter
$d_{\max}$ and parameterized on the requested content $z$, are given by\footnote{Expression~\eqref{eq:Y_OFF_IMM_mean} is provided in terms of the
function $g_{\text{dB}}(d)=10\cdot\log_{10}(g(d))$, as it it is more
suitable to a numeric integration of the last term.} (see Appendix~\ref{sec:Appendix_B})

\begin{align}
F_{Y,\text{NR,off,im}}\left(y;z\right) & =\frac{1}{1-e^{-\rho_{z}2d_{\max}}}\left(1-e^{-\rho_{z}2g_{\text{dB}}^{-1}\left(10\log_{10}\left(\frac{1}{y}\sigma_{c}^{2}\left(2^{\bar{e}}-1\right)\right)\right)}\right)u_{[0,\sigma_{c}^{2}\left(2^{\bar{e}}-1\right)/g(d_{\max})]}\left(y\right).\label{eq:Y_OFF_IMM_CDF}\\
p_{Y,\text{NR,off,im}}\left(y;z\right) & =-\frac{1}{y^{2}}\frac{2\rho_{z}\sigma_{c}^{2}\left(2^{\bar{e}}-1\right)}{\left(1-e^{-\rho_{z}2d_{\max}}\right)}\frac{e^{-\rho_{z}2g^{-1}\left(\frac{1}{y}\sigma_{c}^{2}\left(2^{\bar{e}}-1\right)\right)}}{g'\left(g^{-1}\left(\frac{1}{y}\sigma_{c}^{2}\left(2^{\bar{e}}-1\right)\right)\right)}u_{[0,\sigma\left(2^{\bar{e}}-1\right)/g\left(d_{\max}\right)]}\left(y\right).\label{eq:Y_OFF_IMM_PDF}\\
\overline{Y}_{\text{NR,off,im}}^{(z)}\left(d_{\max}\right) & =\frac{\sigma_{c}^{2}(2^{\bar{e}}-1)}{g(d_{\max})}\left(\frac{1}{1-e^{\rho_{z}2d_{\max}}}\right)+\frac{\sigma_{c}^{2}\left(2^{\bar{e}}-1\right)}{1-e^{-\rho_{z}2d_{\max}}}\frac{\ln10}{10}\int_{g_{\text{dB}}\left(d_{\max}\right)}^{+\infty}\hspace{-1cm}10^{-y'/10}e^{-\rho_{z}2g_{\text{dB}}^{-1}\left(y'\right)}dy'.\label{eq:Y_OFF_IMM_mean}
\end{align}

In the case the request is not fulfilled immediately, but it is fulfilled
within the content timeout, through an encounter with another device,
the transmission distance is always equal to $d_{\max}$ (for any requested content $z$), as we
are assuming that as soon as the two devices get within each other's
range, the content is transmitted. Therefore, the transmit power is,
in this case, see \eqref{eq:tx_power_as_a_function_of_distance}\medskip{}
\begin{equation}
Y_{\text{NR,off,del}}\left(d_{\max}\right)=\sigma_{c}^{2}\left(2^{\bar{e}}-1\right)/g(d_{\max}).\label{eq:Y_OFF_DEL}
\end{equation}\medskip{}

In the case that the request is fulfilled using an I2D transmission,
the transmit power is a function of the distance $d^{(\text{I2D)}}$
between the eNodeB and the receiving device as in\medskip{}
\begin{equation}
Y_{\text{NR,non-off}}\left(d^{(\text{I2D)}}\right)=\sigma_{c}^{2}\left(2^{\bar{e}}-1\right)/g(d^{(\text{I2D)}}),\label{eq:Y_NR_NON_OFF_DEFINITION}
\end{equation}
and, again, this does not depend on the requested content $z$.\medskip{}

The distance $d^{\text{(I2D)}}$ is distributed uniformly in the range
$[0,d_{\max}^{\text{(I2D)}}]$, where $d_{\max}^{\text{(I2D)}}$ is
the cell radius\footnote{We assume that the ROI is fully covered by a set of eNodeBs, each
with coverage $d_{\max}^{\text{(I2D)}}$.}. With some integral calculus, see Appendix B, it can be showed that
the CDF, PDF, and average value of $Y_{\text{NR,non-off}}$ are given
by
\begin{align}
F_{Y,\text{NR,non-off}}\left(y\right)= & \frac{1}{d_{\max}^{\text{(I2D)}}}g_{\text{dB}}^{-1}\left(10\log_{10}\left(\frac{1}{y}\sigma_{c}^{2}\left(2^{\bar{e}}-1\right)\right)\right)u_{[0,\sigma_{c}^{2}\left(2^{\bar{e}}-1\right)/g\left(d_{\max}^{\text{(I2D)}}\right)]}\left(y\right),\label{eq:Y_NON_off_CDF}\\
p_{Y,\text{NR,non-off}}\left(y\right)= & -\frac{1}{y}\frac{1}{d_{\max}^{\text{(I2D)}}}\frac{1}{g_{\text{dB}}'\left(g_{\text{dB}}^{-1}\left(10\log_{10}\left(\frac{1}{y}\sigma_{c}^{2}\left(2^{\bar{e}}-1\right)\right)\right)\right)}\frac{10}{\ln10}u_{[0,\sigma_{c}^{2}\left(2^{\bar{e}}-1\right)/g\left(d_{\max}^{\text{(I2D)}}\right)]}\left(y\right),\label{eq:Y_NON_off_PDF}\\
\overline{Y}_{\text{NR,non-off}}\left(d_{\max}^{\text{(I2D)}}\right)= & \frac{\sigma_{c}^{2}(2^{\bar{e}}-1)}{g(d_{\max}^{\text{(I2D)}})}-\frac{\sigma_{c}^{2}\left(2^{\bar{e}}-1\right)}{d_{\max}^{\text{(I2D)}}}\frac{\ln10}{10}\int_{g_{\text{dB}}\left(d_{\max}^{\text{(I2D)}}\right)}^{+\infty}10^{-y'/10}g_{\text{dB}}^{-1}\left(y'\right)dy'.\label{eq:Y_NON_off_mean}
\end{align}\medskip{}

The overall average transmit power used for fulfilling a non-repeated
request for content $z$ through D2D offloading, expressed as a function
of $d_{\max}$ and $d_{\max}^{\text{(I2D)}}$, $d_{\max}^{\text{(I2D)}}$, and of the specific requested content
$z$, is given by\medskip{}
\begin{align}
\overline{\mathcal{P}}_{tx,\text{off}}\left(d_{\text{\ensuremath{\max}}},z\right)=\, & Pr^{(d_{\max})}\left(\text{off.imm}\mid z\right)\overline{Y}_{\text{off,imm}}\left(d_{\text{\ensuremath{\max}}}\right)\nonumber \\
 & +Pr^{(d_{\max})}\left(\text{off.del}\mid z\right)Y_{\text{off,del}}\left(d_{\text{\ensuremath{\max}}}\right)\nonumber \\
 & +\left(1-Pr^{(d_{\max})}\left(\text{off.imm}\mid z\right)\right.\left.-Pr^{(d_{\max})}\left(\text{off.del}\mid z\right)\right)\overline{Y}_{\text{NR,non-off}}\left(d_{\max}^{\text{(I2D)}}\right).\label{eq:average_TX_pow_param_z}
\end{align}
Note that, in the left-hand side of  \eqref{eq:average_TX_pow_param_z}, the dependence on the $d_{\max}^{\text{(I2D)}}$
is dropped, as we consider it a system parameter.

\noindent Finally, our concluding result is the following
\begin{thm}
Under the assumptions in Subsection~\ref{subsec:Vehicle-arrival-and-content-request-model},
the average power for fulfilling a non-repeated request is given by
\begin{align}
\mathcal{\overline{P}}_{tx}\left(d_{\text{\ensuremath{\max}}}\right)= & \sum_{z=1}^{N_{Z}}p_{Z}\left(z\mid\mathrm{NR}\right)\overline{\mathcal{P}}_{tx}\left(d_{\text{\ensuremath{\max}}},z\right),\label{eq:average_TX_pow}
\end{align}
where $p_{Z}\left(z\mid NR\right)$ is given by~\eqref{eq:Prob_Z_conditioned_NR}
and $\overline{\mathcal{P}}_{tx}\left(d_{\text{\ensuremath{\max}}},z\right)$
is given by \eqref{eq:average_TX_pow_param_z}.
\end{thm}
\begin{IEEEproof}
The average transmit power of a non-repeated request, irrespective
of which content is requested, can be obtained by averaging~\eqref{eq:average_TX_pow_param_z}
over all the possible events $\left(Z=z\mid\text{NR}\right)$ that
the requested content is $z$, conditioned on the fact that the request
is not repeated. The average is obtained assigning a weight $Pr\left(Z=z\mid\text{NR}\right)=p_{Z}\left(z\mid\mathrm{NR}\right)$,
see~\eqref{eq:Prob_Z_conditioned_NR}, to the value of the average
transmit power required to transmit content $z$, and summing the
product over all the possible realizations of $Z$. In this way, the
desired result~\eqref{eq:average_TX_pow} is obtained.
\end{IEEEproof}
The expression of the average transmit power as a function of $d_{\max}$
allows to select this system parameter to minimize the overall system
energy consumption or, formally,
\begin{equation}
d_{\max}^{(\text{opt})}=\underset{d_{\max}\in\mathbb{R}^{+}}{\arg\min}\mathcal{\overline{P}}_{tx}\left(d_{\text{\ensuremath{\max}}}\right).
\end{equation}
Despite the the analytical expressions involved in \eqref{eq:average_TX_pow}
(specifically, \eqref{eq:Y_OFF_IMM_mean}, \eqref{eq:Y_OFF_DEL},
\eqref{eq:Y_NON_off_mean}) make it hard to compute the optimal $d_{\max}$
in closed form, \eqref{eq:average_TX_pow} can be computed through
numerical integration in a relatively straightforward way, under the
assumption of specific models for $p_{Z}\left(z\right)$ and $p_{V}(v)$.
Our results (see Section~\ref{sec:Performance-evaluation}) show
that, as intuition suggests (it can also be proved formally, but we
do not do it here for space reasons), $\mathcal{\overline{P}}_{tx}\left(d_{\text{\ensuremath{\max}}}\right)$
is convex with respect to $d_{\max}$, making it straightforward to
find the $d_{\max}$ which minimizes the average transmit power, and
hence the energy consumption (see footnote \ref{fn:power_energy}).

\section{Performance evaluation\label{sec:Performance-evaluation}}

To evaluate our results, we used a custom simulator written in Matlab.
We used the following settings. The ROI consists of a street chunk
of length 1,8~Km and width 20~m. The distance between the centers
of the two lanes (one lane per marching direction), is 10~m. There
is an eNodeBs every 600~m, located at 0, 600, 1200, and 1800~m,
respectively, from the left edge of the ROI\footnote{To avoid border effects, we only consider content requests fulfilled
when the receiving device is located under the coverage of the two
central eNodeBs.}. Vehicles enter the street with an arrival rate of $\lambda_{t}=1/3$
(one vehicle every 3 seconds). The vehicles speed is distributed uniformly
in the range $[v_{a},v_{b}]$, with $v_{a}=6\,\text{m/s}$ and $v_{b}=16\,\text{m/s}$.
Each user issues content requests at a rate of 10 requests per minute
(including repeated requests). The content library $\mathcal{L}$
has size $N_{Z}=10^{4}$ contents, and the PMF representing the content
popularity is a Zipf distribution, i.e $p_{Z}(z)\sim\frac{1}{\zeta(\alpha)}z^{-\alpha}$,
truncated at the value of the library size and with $\alpha=1.1$.
We have assumed contents of equal size of 500~KBytes.

Our simulator reproduces a MAC/physical layer which allows concurrent
D2D and/or I2D transmissions to be possibly allocated the same portions
of spectrum, provided that the corresponding transmitter-receiver
pairs are sufficiently far apart. More specifically, in our implementation,
we have used a slightly modified version of the presented in \cite{Yang2017}.
For space reasons, we do not provide full details of our implementation.
Suffice it to say that, differently from \cite{Yang2017}, we do allow
for different transmit power selection in different links, as we assume
that transmit power is a function of the transmitter-receiver distance
through Eq.~\eqref{eq:tx_power_as_a_function_of_distance}.

Time is organized in frames, and in each frame both D2D and I2D communications
can be scheduled, as a result of the decisions of the CDMS described
in Subsection~\ref{subsec:Content-Dissemination-Management}. We
have assumed contents of equal size equal to 500~KBytes. The nominal
channel model $g(d)$ is the one provided by Equations~5-4~and~5.5
of \cite{METIS_chanmod}\footnote{The models in \cite{METIS_chanmod} are based on large scale measurements
campaigns conducted in collaboration with large Telecom companies
like Nokia and Docomo. As such, it represents a state of the art reference
channel models for a variety of scenarios.}. The system bandwidth is 10~MHz. The central carrier frequency is
$2.3\,\mathrm{GHz}$. Content deliveries are scheduled by the CDMS
every second, and the radio resource allocation scheduler (see Subsection~\ref{subsec:Energy-consumption})
allocates PRBs of width 200~KHz and duration $1\,\text{ms}$. In
each scheduling period, the overall number of PRBs that can be assigned
is $50000$. In each 200~KHz block there are 12 equally spaced subcarriers.
The target normalized nominal information rate $\bar{e}$ is set to
5~bps/Hz. Accordingly, a PRB carries 1~Kbit, and each content transmission
requires 400~PRBs. The transmit power is selected according to~\eqref{eq:tx_power_as_a_function_of_distance},
with $\sigma_{c}^{2}=-164\,\mathrm{dBm/Hz}$, plus a link margin of
15 dB\footnote{The value selected for $\sigma_{c}^{2}$ stems from adding a typical
noise figure of $F_{rc}=10\,\mathrm{dB}$ to the typical thermal noise
spectral density $N_{0,\text{[dBm]/Hz}}=-174\,\mathrm{dBm/Hz}$. }.

We considered 29 equally spaced values for the maximum D2D transmission
distance $d_{\max}$ in the range {[}20,300{]} m. For each value of
$d_{\max}$, we run 10 independent i.i.d simulations, each lasting
15 minutes, reinitializing the random number generator seed with
the same state at the beginning of each batch of 10 simulations. Each
simulation is initialized with a random number of vehicles, position
and speed of each vehicle, according to the assumptions and the preliminary
results in Subsections~\ref{subsec:Vehicle-arrival-and-content-request-model}
and \ref{subsec:Preliminary-results}, respectively. The content cache
of each node is initialized according to~\eqref{eq:Prob_content_z_in_cache-1}.
The content timeout is set to $\tau_{c}=20\,$~s, and the sharing
timeout to $\tau_{s}=600$~s.

\begin{figure}[t]
\begin{centering}
\includegraphics[width=0.7\columnwidth]{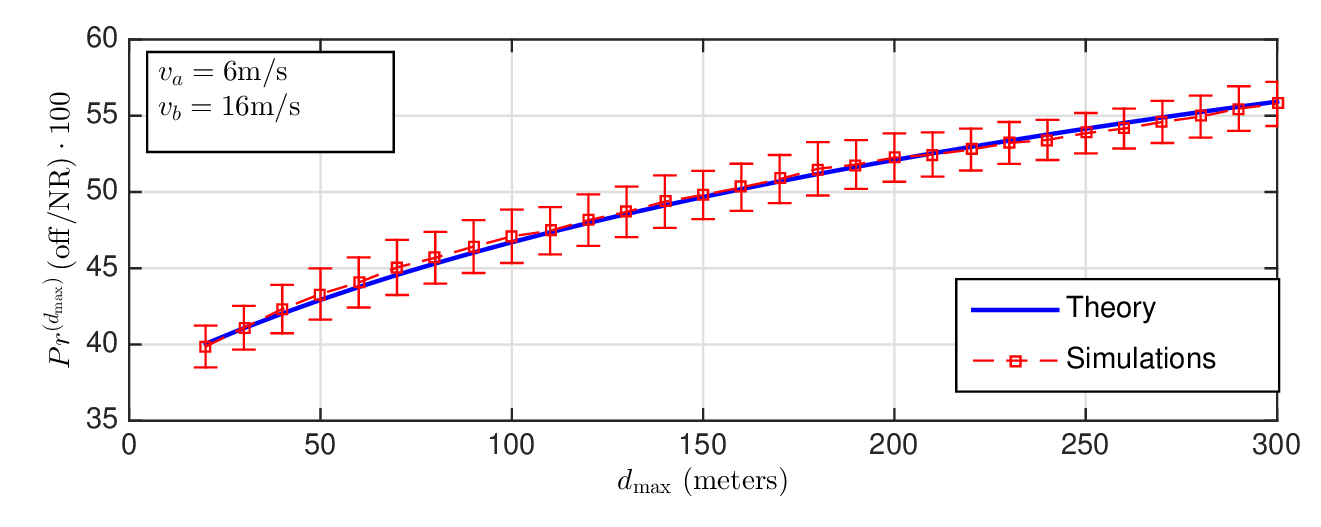}\vspace{-3mm}
\par\end{centering}
\centering{}\caption{Offloading efficiency}
\label{fig:eff}
\end{figure}

\begin{figure}[t]
\begin{centering}
\includegraphics[width=0.7\columnwidth]{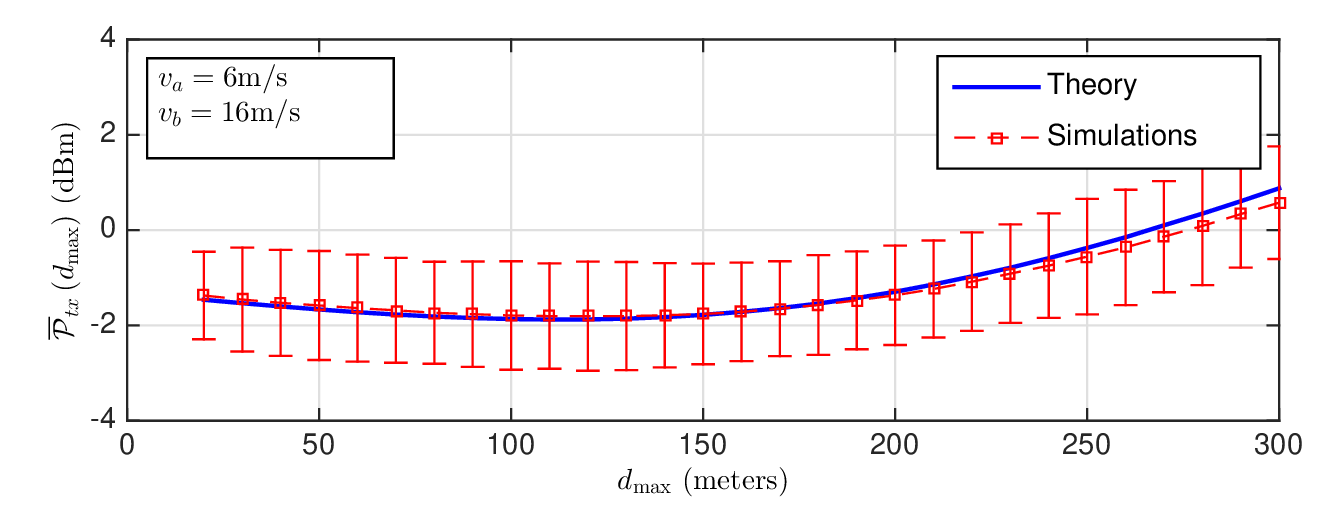}\vspace{-3mm}
\par\end{centering}
\centering{}\caption{System-wise average transmit power per subcarrier}
\label{fig:pow}
\end{figure}

Figures~\ref{fig:eff}~and~\ref{fig:pow} represent the average
performance obtained in the simulations (with 95\% confidence intervals)
and the performance predicted by our model, in terms of offloading
efficiency and average required transmit power per subcarrier. It
can be seen that the theoretical results provide a perfect match of
the performance obtained through simulation, despite our theoretical
model overlooks some details, for instance through the unidimensional
representation in the model of the ROI (in simulations, we did reproduce
two different lanes corresponding to the opposite marching directions).
As intuition suggests, the offloading efficiency increases with the
maximum transmission range of the devices, since the probability that
within a distance $d_{\max}$ there is a neighbor with the desired
content, increases. However, keeping increasing the distance indefinitely,
does not help in terms of power consumption, as the transmit power
to reach ``far'' neighbors increases exponentially. Moreover, keeping
increasing $d_{\max}$, the marginal gain in terms of offloading efficiency
(represented by the slope of the offloading efficiency curve), progressively
diminishes. On the other side, decreasing $d_{\max}$ reduces the
probability to find neighbors, and hence, for low values of $d_{\max}$,
the transmit power is dominated by the term related to I2D transmissions,
which may require transmissions at a distance larger than $d_{\max}$
(up to 300 m in our example). As a result, there is an optimal value
for the maximum distance, whose selection guarantees the minimization
of the average transmit power, and therefore of the overall system
energy consumption, which is related to the average transmit power
through a constant term.

\section{Conclusion\label{sec:Conclusion}}

We have considered a D2D data offloading content delivery system for
a mobile environment, and specifically a vehicular scenario. We have
derived an analytical model based on results of the theory of point
processes (in this case unidimensional) to compute the average system
offloading probability (or equivalently, offloading efficiency) and
average transmit power (or equivalently, the system wise energy consumption).
The derived expressions are function of different parameters related
to various domains, namely, to the content request process (content
interest distribution and requests arrival rate), the vehicles mobility
model and density, to the path loss model, and to the range within
which devices should be considered as neighbors, or, equivalently,
the maximum power at which devices should be allowed to transmit.
We have checked that the proposed model, although obtained by overlooking
several details of the system geometry, is able to predict in an accurate
way the optimal value for the devices transmission range or, equivalently,
the maximum transmit power that should be allowed to the devices.

\section*{Acknowledgment}

This work was partially funded by the EC under the H2020 REPLICATE
(691735), SoBigData (654024) and AUTOWARE (723909) projects.
\begin{center}
\bibliographystyle{IEEEtran}

\par\end{center}

\appendices{}

\section{Proof of Lemmas 1-3}

\subsubsection*{Proof of Lemma~\ref{lem:1}}

Under the assumption that the speeds of the vehicles are statistically
independent, the fraction of vehicles with speed in the range $[v,v+\Delta v]$,
for a finite $\Delta v$, is equal to the probability that the speed
of a vehicle entering the ROI is in the range $[v,v+\Delta v]$, i.e.,
$Pr\left(V\in[v,v+\Delta v]\right)=\int_{v}^{v+\Delta v}p_{V}(v')dv'$
. The process of arrival of vehicles with speed in the same range
corresponds to the overall homogeneous TPPP of vehicles arrivals,
thinned with the probability $Pr\left(V\in[v,v+\Delta v]\right)$.
Accordingly, the arrival rate of the thinned TPPP is $\lambda_{t,[v,v+\Delta v]}=\lambda_{t}\int_{v}^{v+\Delta v}p_{V}(v')dv'.$
This expression allows us to write the differential of the (overall)
arrival rate $\lambda_{t}$ as a function of the differential of the
speed $dv$. Specifically\footnote{We use the identity $x^{*}=\int_{0}^{x^{*}}dx$, that holds for any
Real quantity~$x^{*}$. }
\begin{equation}
d\lambda_{t}=\lambda_{t}p_{V}(v)dv.\label{eq:differential_temporal_arrival_rate-1}
\end{equation}
This can be considered as the arrival rate of the elementary process
of arrival of vehicles with speed equal to an exact value $v$.

In one second, a vehicle traverses a street span of $\left|v\right|\cdot1$
meters. In the meantime, on average, $d\lambda_{t}$ vehicles have
entered the same street span of length $\left|v\right|\cdot1$ meters.
Transforming the time elapsed from entering the ROI into the distance
spanned, since the speed is constant and the arrival instants are
independent, it can be easily showed that the \emph{spatial} distribution
of such vehicles is a homogeneous SPPP with density, obtained using~(\ref{eq:differential_temporal_arrival_rate-1}),
\begin{equation}
\small{d\rho=\frac{1}{\left|v\right|}d\lambda_{t}=\frac{1}{\left|v\right|}\lambda_{t}p_{V}(v)dv.}\label{eq:differential_spatial_(linear)_density-1}
\end{equation}

The overall spatial point process of vehicles present on the street,
at any given instant, is the superposition of an infinite number of
elementary homogeneous SPPPs of the kind above (one for each value
of $v$). Therefore, it is again a homogeneous SPPP with linear density
given by\footnote{We use the transformation of the integration variable from $\rho$
to $v$ using the relation~(\ref{eq:differential_spatial_(linear)_density-1})
between their differentials.} $\rho=\int_{0}^{\rho}d\rho'=\int_{-\infty}^{+\infty}\frac{1}{\left|v\right|}\lambda_{t}p_{V}(v)dv$,
which coincides with~\eqref{eq:overall_spatial_density}. 

Using~\eqref{eq:PDF_v_special} in~\eqref{eq:overall_spatial_density}
we obtain
\begin{align*}
\small{\rho} ~& \small{=\frac{\lambda_{t}}{2(v_{b}-v_{a})}\int_{-\infty}^{+\infty}\frac{1}{\left|v\right|}\left(u_{[-v_{b},-v_{a}]}(v)+u_{[v_{a},v_{b}]}(v)\right)dv}\\
 & \small{=\frac{\lambda_{t}}{2(v_{b}-v_{a})}2\int_{v_{a}}^{v_{b}}\frac{1}{v}dv=\frac{\lambda_{t}(\ln v_{b}-\ln v_{a})}{(v_{b}-v_{a})},}
\end{align*}
i.e, \eqref{eq:overall_spatial_density_special}. 

Let $x(t)$ be the trajectory of a vehicle moving at speed $v*$,
and $x'(t)=x(t)+d_{\max}$ the trajectory of a point displaced at
a distance $d_{\max}$ from it. The time process of the instants
at which $x'(t)$ coincides with the position of other vehicles, \emph{that
have a specific speed} $v$, is determined by the relative speed $\left(v-v^{*}\right)$.
Consider a reference system which moves across space and has origin,
at each instant, in $x'(t)$. In this reference system, the speed
of the considered vehicles is $\left(v-v^{*}\right)$. The vehicles
``encountered'' by the initially considered vehicle moving at speed
$v^{*}$, during an interval of 1 second starting at a given instant
$t_{0}$, are those that, at instant $t_{0}$, are positioned (in
the new reference system) along the segment originating at the position
$x_{a}=0$ (the position of the displaced point in the new reference
system) and the position $x_{b}=\left(v-v^{*}\right)$. Since the
positions of the vehicles at $t_{0}$ is a homogeneous SPPP (with
density $d\lambda_{s}$ given by \eqref{eq:differential_spatial_(linear)_density-1})
and $\left(v-v^{*}\right)$ is constant, transforming travelled distances
into time intervals we can claim that the set of the instants at which
the devices are encountered is a homogeneous TPPP. We call the rate
of this process ``elementary encountering rate'', and indicate it
with $d\lambda_{e}^{*}$. We can compute $d\lambda_{e}^{*}$ using
\eqref{eq:differential_spatial_(linear)_density-1}\footnote{Essentially, \eqref{eq:differential_spatial_(linear)_density-1} expresses
the relation between spatial and temporal elementary intensities (i.e.,
a linear density and a rate) of the two elementary PPPs we are considering:
the SPPP of the positions of vehicles traveling at a specific speed
$v$, and the TPPP of the instant they cross a given point.}, by replacing $d\lambda_{t}$ with $d\lambda_{e}^{*}$, obtaining\footnote{In \eqref{eq:differential_encountering_rate}, we use the modulus
to let all the possible values of speeds $v^{*}$ and $v$ result
in positive encountering rates. These possibilities encompass having
vehicles traveling in the same  or opposite directions.}
\begin{equation}
\small{d\lambda_{e}^{*}=\left|v^{*}-v\right|d\rho=\frac{\left|v^{*}-v\right|}{\left|v\right|}\lambda_{t}p_{V}(v)dv,}\label{eq:differential_encountering_rate}
\end{equation}

Following the same line of reasoning used above, the temporal process
of vehicles moving at \emph{any} speed encountered by a vehicle moving
at speed $v^{*}$ is the superposition of an infinite number of elementary
``encountering processes'' of the kind above. Accordingly, the rate
at which the a point moving at speed $v^{*}$ encounters other vehicles
at \emph{any} speed is 
\begin{equation}
\small{\lambda_{e}^{*}=\int_{-\infty}^{\infty}\lambda_{t}p_{V}(v)\frac{\left|v^{*}-v\right|}{\left|v\right|}dv.}\label{eq:overall_device_encountering_rate_with_general_speed_PDF-1}
\end{equation}
Thus, we have obtained~\eqref{eq:overall_device_encountering_rate_with_general_speed_PDF}.
Finally, using~\eqref{eq:PDF_v_special} in~\eqref{eq:overall_device_encountering_rate_with_general_speed_PDF-1},
with a few integral calculus steps, it is easy to obtain~\eqref{eq:overall_device_encountering_rate_with_general_speed_PDF_SPECIAL}.

\subsubsection*{Proof of Lemma~\ref{lem:2}}

Result~\eqref{eq:arrival-rate-of-TPPP-req_for_content_z-1} can be
obtained as follows: since the content request arrival process of
each device is a homogeneous TPPP, and the requests are statistically
independent, the process of issuing requests for a specific content
$z$ by a given device is again a homogeneous TPPP, which results
from thinning the TPPP of the overall requests issued by a device,
with the probability that the requested content is $z$, i.e., $p_{Z}\left(z\right)$.
The arrival rate of the thinned process, $\lambda_{z}$, is the product
of the arrival rate of the content request arrival process of each
device, $\lambda_{Z}$, by the same probability, or $p_{Z}\left(z\right)$,
or $\lambda_{z}=p_{Z}(z)\lambda_{Z}$, i.e., \eqref{eq:arrival-rate-of-TPPP-req_for_content_z-1}.

Result~\eqref{eq:Prob_content_z_in_cache-1} provides upper and lower
bounds to the probability $Pr\left(\mathcal{C}\ni z\right)$ that
at a given instant $t$, the cache $\mathcal{C}$ of a generic device
contains a specific content $z$. We observe that a device cache holds
a content $z$ at a given instant $t$ if the device has previously
requested the content, has obtained it, and has not yet removed it
from its cache (due to the expiration of the sharing timeout $\tau_{s}$).
The probability of this event is lower bounded by the probability
that the device has requested the content (one or more times) in the
interval $[t-\tau_{s},t-\tau_{c}]$ and upper bounded by the probability
that the device has requested the content (one or more times) in the
interval $[t-\tau_{s},t]$. Since the content request process for
content $z$ by a generic device is a TPPP with rate given by (\ref{eq:arrival-rate-of-TPPP-req_for_content_z-1}),
the number of content requests in a given interval is a Poisson random
variable with parameter equal to the product of $\lambda_{z}$ times
the duration of the interval. Accordingly, $Pr\left(\mathcal{C}\ni z\right)$
has the upper and lower bounds in~\eqref{eq:Prob_content_z_in_cache-1}.

Result~\eqref{eq:SPPP-geo-content-distribution-2} can be obtained
building on the SPPP with density $\rho$ of the devices' positions
at a given instant. Since the content requests are independent across
devices, the content of their caches at a given instant are approximately
independent\footnote{There is a small correlation among the contents present in the devices'
caches at a given instant $t$. This is related to the presence of
the contents requested in the interval $[t-\tau_{c},t]$, whose reception
depends on the composition of the surrounding nodes' caches during
that interval. Compared to the amount of contents received in the
interval $[t,t-\tau_{c}]$, which are certainly in the devices' caches,
since $\tau_{s}\gg\tau_{c}$, the former set of contents has a minimal
weight in the overall statistics of the caches. The latter sets of
contents are certainly independent since they have been received,
irrespective of the devices trajectories and caches during the interval
$[t-\tau_{s},t]$.}. The presence of content $z$ in the caches of devices located in
the ROI is a again a SPPP of the same type, but with density given
by $\rho_{z}=Pr\left(\mathcal{C}\ni z\right)\rho$, i.e, \eqref{eq:SPPP-geo-content-distribution-2}.

Result~\eqref{eq:encountering_rate_with_devices_with_content_z-1}
provides the rate at which a point moving at speed $v^{*}$ encounters
vehicles with a specific content $z$ in their caches. In Subsection~\ref{subsec:Preliminary-results}
(Lemma~\ref{lem:1}), we showed that the process of encountering
devices, for a point moving at speed $v^{*}$, is a homogeneous TPPP
with rate $\lambda_{e}^{(v^{*})}$. Due to the (approximate) independence
of the caches contents, the process of encountering devices that have
a specific content $z$ in their caches is still a homogeneous TPPP
resulting from thinning the overall encountering TPPP with the probability
$Pr\left(\mathcal{C}\ni z\right)$ that content $z$ is cached at
the encountered devices. The corresponding encountering rate is given
by the product of the overall encountering rate (\ref{eq:overall_device_encountering_rate_with_general_speed_PDF})
times $Pr\left(\mathcal{C}\ni z\right)$, or $\lambda_{e,z}^{(v^{*})}=Pr\left(\mathcal{C}\ni z\right)\lambda_{e}^{(v^{*})},$
i.e., \eqref{eq:encountering_rate_with_devices_with_content_z-1},
with $\lambda_{e}^{(v^{*})}$ given by \eqref{eq:overall_device_encountering_rate_with_general_speed_PDF}.

\subsubsection*{Proof of Lemma~\ref{lem:3}}

Consider a device whose cache content is $\mathcal{C}$ which issues
a content request. The event that the request is non-repeated, prior
to the realization of the specific requested content, is the union
of the (overlapping) events $\left(\mathcal{C}\not\ni z\right),\,\forall z\in\mathcal{L}$.
By the law of total probability, we have
\begin{equation}
\small{Pr\left(\mathrm{NR}\right)=\sum_{z}Pr\left(Z=z\right)Pr\left(\mathcal{C}\not\ni z/Z=z\right).}\label{eq:laux_eq_1}
\end{equation}

Under the assumption on the content request process in Subsection~\ref{subsec:Vehicle-arrival-and-content-request-model},
the random variable $Z$ representing the requested content, and the
set $\mathcal{C}$ of the contents in the cache at the time of request,
are statistically independent. Therefore, we can write $Pr\left(\mathcal{C}\not\ni z\mid Z=z\right)=Pr\left(\mathcal{C}\not\ni z\right)$,
which plugged in~\eqref{eq:laux_eq_1}, gives the desired result~\eqref{eq:Prob_non_repeated_req}. 

Now, the probability that the requested content is $z$, conditioned
on the fact that the request is not repeated, is
\begin{equation}
\small{p_{Z}\left(z\mid\text{NR}\right)\triangleq Pr\left(Z=z\mid\text{NR}\right)=\frac{Pr\left(Z=z,\text{NR}\right)}{Pr\left(\text{NR}\right)}.}\label{eq:aux_eq_2}
\end{equation}

The joint probability $Pr\left(Z=z,\text{NR}\right)$ can be written,
by the Bayes Theorem, as $Pr\left(Z=z,\text{NR}\right)=Pr\left(\text{NR}\mid Z=z\right)Pr(Z=z)$.
But $Pr\left(\text{NR}\mid Z=z\right)$ is the probability
that the cache of the requesting node does not contain the specific
content $z$ (once it has been determined), i.e., $Pr\left(\mathcal{C}\not\ni z\right)$.
Therefore, we can replace the numerator in ~\eqref{eq:aux_eq_2}
with the product $Pr\left(Z=z\right)Pr\left(\mathcal{C}\not\ni z\right)$.
Finally, replacing the denominator in~\eqref{eq:aux_eq_2} with~\eqref{eq:laux_eq_1},
we obtain the desired result~\eqref{eq:Prob_non_repeated_req}.

\section{Proof of the results in Subsection~\ref{subsec:Energy-consumption}\label{sec:Appendix_B}}

\subsection{Cumulative Distribution Function, Probability Density Function, and
average value of the transmit power used for immediate content delivery
through D2D.}

In the following, we provide the steps to obtain expressions \eqref{eq:Y_OFF_IMM_CDF},
\eqref{eq:Y_OFF_IMM_PDF}, and \eqref{eq:Y_OFF_IMM_mean}. The CDF
of $Y_{\text{NR,off,im}}$ in \eqref{eq:Y_OFF_IMM_CDF} can be computed
as
\begin{align}
F_{Y,\text{NR,off,im}}\left(y\middle|D\leq d_{\max}\right) & =Pr\left(Y_{\text{NR,off,im}}\leq y\middle|D\leq d_{\max}\right)\nonumber\\
 & =Pr\left(\frac{1}{g(D)}\sigma_{c}^{2}\left(2^{\bar{e}}-1\right)\leq y\middle|D\leq d_{\max}\right)\nonumber\\
\hphantom{F_{Y,\text{off,im}}\left(y\right)} & =Pr\left(g(D)\geq\frac{1}{y}\sigma_{c}^{2}\left(2^{\bar{e}}-1\right)\middle|D\leq d_{\max}\right)\nonumber\\
 & =Pr\left(g_{\text{dB}}(D)\geq10\log_{10}\left(\frac{1}{y}\sigma_{c}^{2}\left(2^{\bar{e}}-1\right)\right)\middle|D\leq d_{\max}\right)\nonumber\\
 & =Pr\left(D\leq g_{\text{dB}}^{-1}\left(10\log_{10}\left(\frac{1}{y}\sigma_{c}^{2}\left(2^{\bar{e}}-1\right)\right)\right)\middle|D\leq d_{\max}\right).
\end{align}
i.e, it is given by the distribution function of the distance among
a point of the SPPP process and its closest neighbor, conditioned to
the distance being less than $d_{\max}$, and evaluated at $g_{\text{dB}}^{-1}\left(10\log_{10}\left(\frac{1}{y}\sigma_{c}^{2}\left(2^{\bar{e}}-1\right)\right)\right)$.
The conditional distribution $F_{D}\left(d\middle|D\leq d_{max}\right)\triangleq Pr\left(D\leq d\middle|D\leq d_{max}\right)=\frac{Pr\left(D\leq d,D\leq d_{max}\right)}{Pr\left(D\leq d_{max}\right)}$,
for the distances $d\leq d_{\max}$ of interest\footnote{By construction, $g^{-1}\left(\frac{1}{y}\sigma_{c}^{2}\left(2^{\bar{e}}-1\right)\right)$
is certainly less than $d_{\max}$.}, coincides with $Pr\left(D\leq d\right)/Pr\left(D\leq d_{max}\right).$
Using the expression of the CDF of the nearest neighbor distance for
homogeneous unidimensional SPPPs $F_{D}\left(d\right)=1-e^{-\rho_{z}2d}$,
we obtain $F_{D}\left(d\middle|D\leq d_{max}\right)=u_{[0,d_{\max}]}(d)\left(1-e^{-\rho_{z}2d}\right)/\left(1-e^{-\rho_{z}2d_{\max}}\right)$
and, ultimately,
\begin{align}
F_{Y,\text{NR,off,im}}\left(y;z\right) & =\frac{1-e^{-\rho_{z}2g_{\text{dB}}^{-1}\left(10\log_{10}\left(\frac{1}{y}\sigma_{c}^{2}\left(2^{\bar{e}}-1\right)\right)\right)}}{1-e^{-\rho_{z}2d_{\max}}}u_{[0,\sigma_c^2\left(2^{\bar{e}}-1\right)/g\left(d_{\max}\right)]}\left(y\right)\label{eq:appendixBenergyCDFoffimm},
\end{align}
where, in the left-hand side, we have made explicit the dependence on the requested content $z$, as the density $\rho_z$ now appears in the right-hand side, and removed the conditioning event $\left(D\leq d_{\max};z\right)$ , as the transmission distance is less than $d_{\max}$ by construction. Eq. \eqref{eq:appendixBenergyCDFoffimm} is the desired result, i.e., Eq.~\eqref{eq:Y_OFF_IMM_CDF}.

For the sake of completeness, it is worth also computing the PDF of
$Y_{\text{NR,off,im}}$, even though it is not required to compute
its average value, since the integration by parts in the computation
of the average value gets rid of the PDF (see below). The PDF of $Y_{\text{NR,off,im}}$
is given by the first derivative of the CDF, i.e.:
\begin{align}
p_{Y,\text{NR,off,im}}(y;z) ~ =&~ \frac{d}{dy}F_{Y,\text{off,im}}\left(y;z\right)\nonumber\\
 =&~\frac{d}{dy}\left(\frac{1-e^{-\rho_{z}2g_{\text{dB}}^{-1}\left(10\log_{10}\left(\frac{1}{y}\sigma_{c}^{2}\left(2^{\bar{e}}-1\right)\right)\right)}}{1-e^{-2\rho_{z}d_{\max}}}u_{[0,\sigma_c^2\left(2^{\bar{e}}-1\right)/g\left(d_{\max}\right)]}\left(y\right)\right)\nonumber\\
 =&~\frac{1}{1-e^{-2\rho_{z}d_{\max}}}\frac{d}{dy}\left(1-e^{-\rho_{z}2g^{-1}\left(\frac{1}{y}\sigma_{c}^{2}\left(2^{\bar{e}}-1\right)\right)}\right)u_{[0,\sigma_c^2\left(2^{\bar{e}}-1\right)/g\left(d_{\max}\right)]}\left(y\right)\nonumber\\
 =&~\frac{1}{1-e^{-2\rho_{z}d_{\max}}}\frac{d}{dy}\left(\rho_{z}2g^{-1}\left(\frac{1}{y}\sigma_{c}^{2}\left(2^{\bar{e}}-1\right)\right)\right)e^{-\rho_{z}2g^{-1}\left(\frac{1}{y}\sigma_{c}^{2}\left(2^{\bar{e}}-1\right)\right)}\nonumber\\
 &\cdot u_{[0,\sigma_c^2\left(2^{\bar{e}}-1\right)/g\left(d_{\max}\right)]}\left(y\right)\nonumber\\
 =&~\frac{2\rho_{z}}{1-e^{-\rho_{z}2d_{\max}}}\frac{d}{dy}\left(g^{-1}\left(\frac{1}{y}\sigma_{c}^{2}\left(2^{\bar{e}}-1\right)\right)\right)e^{-\rho_{z}2g^{-1}\left(\frac{1}{y}\sigma_{c}^{2}\left(2^{\bar{e}}-1\right)\right)}\nonumber\\
 & \cdot u_{[0,\sigma_c^2\left(2^{\bar{e}}-1\right)/g\left(d_{\max}\right)]}\left(y\right).
\end{align}\vfill

Applying the rule for derivative of an inverse function $\frac{d}{dx}f^{-1}\left(x\right)=\frac{1}{f'\left(f^{-1}\left(x\right)\right)}$,
and the chain rule for the derivative of nested functions $\frac{d}{dx}f\left(g\left(x\right)\right)=f'\left(g\left(x\right)\right)g'\left(x\right)$,
we obtain
\begin{align}
p_{Y,\text{NR,off,im}}\left(y;z\right)= & \frac{2\rho_{z}}{1-e^{-\rho_{z}2d_{\max}}}\frac{1}{g'\left(g^{-1}\left(\frac{1}{y}\sigma_{c}^{2}\left(2^{\bar{e}}-1\right)\right)\right)}\frac{d}{dy}\left(\frac{1}{y}\sigma_{c}^{2}\left(2^{\bar{e}}-1\right)\right)e^{-\rho_{z}2g^{-1}\left(\frac{1}{y}\sigma_{c}^{2}\left(2^{\bar{e}}-1\right)\right)}\nonumber\\
 & \cdot u_{[0,\sigma_{c}^{2}\left(2^{\bar{e}}-1\right)/g\left(d_{\max}\right)]}\nonumber\\
= & \frac{2\rho_{z}}{1-e^{-\rho_{z}2d_{\max}}}\frac{1}{g'\left(g^{-1}\left(\frac{1}{y}\sigma_{c}^{2}\left(2^{\bar{e}}-1\right)\right)\right)}\frac{-\sigma_{c}^{2}\left(2^{\bar{e}}-1\right)}{y^{2}}e^{-\rho_{z}2g^{-1}\left(\frac{1}{y}\sigma_{c}^{2}\left(2^{\bar{e}}-1\right)\right)}\nonumber\\
 & \cdot u_{[0,\sigma_{c}^{2}\left(2^{\bar{e}}-1\right)/g\left(d_{\max}\right)]}\nonumber\\
= & -\frac{1}{y^{2}}\frac{2\rho_{z}\sigma_{c}^{2}\left(2^{\bar{e}}-1\right)}{\left(1-e^{-\rho_{z}2d_{\max}}\right)}\frac{e^{-\rho_{z}2g^{-1}\left(\frac{1}{y}\sigma_{c}^{2}\left(2^{\bar{e}}-1\right)\right)}}{g'\left(g^{-1}\left(\frac{1}{y}\sigma_{c}^{2}\left(2^{\bar{e}}-1\right)\right)\right)}u_{[0,\sigma_{c}^{2}\left(2^{\bar{e}}-1\right)/g\left(d_{\max}\right)]}.
\end{align}\vfill
\noindent Note that this expression is always a non-negative quantity, since
$g'\left(\cdot\right)$ is always negative, because the channel gain
is a decreasing function of its argument.\newpage

The average value of $Y_{\text{off,im}}$ in can be computed as follows

\begin{align}
\overline{Y}_{\text{NR,off,im}}^{(z)}\left(d_{\max}\right)= & \int_{0}^{\sigma_{c}^{2}(2^{\bar{e}}-1)/g(d_{\max})}y\cdot p_{Y,\text{off,im}}\left(y;z\right)dy\nonumber \\
= & \left[y\cdot F_{Y,\text{NR,off,im}}\left(y;z\right)\right]_{0}^{\frac{\sigma_{c}^{2}(2^{\bar{e}}-1)}{g(d_{\max})}}-\int_{0}^{\sigma_{c}^{2}(2^{\bar{e}}-1)/g(d_{\max})}\hspace{-20mm}F_{Y,\text{NR,off,im}}\left(y;z\right)dy\nonumber \\
= & \frac{\sigma_{c}^{2}(2^{\bar{e}}-1)}{g(d_{\max})}\frac{\left(1-e^{-\rho_{z}2g_{\text{dB}}^{-1}\left(10\log_{10}\left(\frac{g(d_{\max})}{\sigma_{c}^{2}(2^{\bar{e}}-1)}\sigma_{c}^{2}\left(2^{\bar{e}}-1\right)\right)\right)}\right)}{1-e^{-\rho_{z}2d_{\max}}}\nonumber \\
 & -\frac{1}{1-e^{-\rho_{z}2d_{\max}}}\int_{0}^{\sigma_{c}^{2}(2^{\bar{e}}-1)/g(d_{\max})}\left(1-e^{-\rho_{z}2g_{\text{dB}}^{-1}\left(10\log_{10}\left(\frac{1}{y}\sigma_{c}^{2}\left(2^{\bar{e}}-1\right)\right)\right)}\right)dy\nonumber \\
= & \frac{1}{1-e^{-\rho_{z}2d_{\max}}}\frac{\sigma_{c}^{2}(2^{\bar{e}}-1)}{g(d_{\max})}\left(1-e^{-\rho_{z}2g_{\text{dB}}^{-1}\left(g_{\text{dB}}(d_{\max})\right)}\right)\nonumber \\
 & -\frac{1}{1-e^{-\rho_{z}2d_{\max}}}\int_{0}^{\sigma_{c}^{2}(2^{\bar{e}}-1)/g(d_{\max})}\left(1-e^{-\rho_{z}2g_{\text{dB}}^{-1}\left(10\log_{10}\left(\frac{1}{y}\sigma_{c}^{2}\left(2^{\bar{e}}-1\right)\right)\right)}\right)dy\label{eq:app2}\\
= & \frac{1}{1-e^{-\rho_{z}2d_{\max}}}\frac{\sigma_{c}^{2}(2^{\bar{e}}-1)}{g(d_{\max})}\left(1-e^{-\rho_{z}2d_{\max})}\right)-\frac{1}{1-e^{-\rho_{z}2d_{\max}}}\int_{0}^{\sigma_{c}^{2}(2^{\bar{e}}-1)/g(d_{\max})}\hspace{-1cm}\hfill1\cdot dy\nonumber \\
 & +\frac{1}{1-e^{-\rho_{z}2d_{\max}}}\int_{0}^{\sigma_{c}^{2}(2^{\bar{e}}-1)/g(d_{\max})}e^{-\rho_{z}2g_{\text{dB}}^{-1}\left(10\log_{10}\left(\frac{1}{y}\sigma_{c}^{2}\left(2^{\bar{e}}-1\right)\right)\right)}dy\label{eq:app3}
\end{align}
We now apply a change of the integration variable $y$ to the variable
\begin{align}
y'\left(y\right) & \triangleq10\log_{10}\left(\frac{1}{y}\sigma_{c}^{2}\left(2^{\bar{e}}-1\right)\right),\label{eq:varchange}
\end{align}
which entails the inverse relation
\begin{align*}
y & =\frac{\sigma_{c}^{2}\left(2^{\bar{e}}-1\right)}{10^{y'/10}}.
\end{align*}
The derivative of $y'$ with respect to $y$ is
\begin{align*}
\frac{d}{dy}y'(y) & =\frac{10}{\ln10}\frac{d}{dy}\ln\left(\frac{1}{y}\sigma_{c}^{2}\left(2^{\bar{e}}-1\right)\right)\\
 & =\frac{10}{\ln10}\frac{1}{\frac{1}{y}\sigma_{c}^{2}\left(2^{\bar{e}}-1\right)}\frac{d}{dy}\left(\frac{1}{y}\sigma_{c}^{2}\left(2^{\bar{e}}-1\right)\right)\\
 & =\frac{10}{\ln10}\frac{y}{\sigma_{c}^{2}\left(2^{\bar{e}}-1\right)}\frac{d}{dy}\left(\frac{1}{y/\sigma_{c}^{2}\left(2^{\bar{e}}-1\right)}\right)\\
 & =\frac{10}{\ln10}\frac{y}{\sigma_{c}^{2}\left(2^{\bar{e}}-1\right)}\left(-\frac{1}{\left(y/\sigma_{c}^{2}\left(2^{\bar{e}}-1\right)\right)^{2}}\right)\frac{d}{dy}\frac{y}{\sigma_{c}^{2}\left(2^{\bar{e}}-1\right)}\\
 & =-\frac{10}{\ln10}\frac{y}{\sigma_{c}^{2}\left(2^{\bar{e}}-1\right)}\left(\frac{\sigma_{c}^{2}\left(2^{\bar{e}}-1\right)}{y}\right)^{2}\frac{1}{\sigma_{c}^{2}\left(2^{\bar{e}}-1\right)}\\
 & =-\frac{10}{\ln10}\frac{\sigma_{c}^{2}\left(2^{\bar{e}}-1\right)}{y}\frac{1}{\sigma_{c}^{2}\left(2^{\bar{e}}-1\right)}\\
 & =-\frac{1}{y}\frac{10}{\ln10}
\end{align*}
The upper and lower integration extremes of the integral appearing
in \eqref{eq:app3}, in terms of $y'$, are
\begin{align*}
y'\left(\frac{\sigma_{c}^{2}(2^{\bar{e}}-1)}{g\left(d_{\max}\right)}\right) & =g_{\text{dB}}\left(d_{\max}\right)\\
y'\left(0\right) & =+\infty.
\end{align*}
So, \eqref{eq:app2} becomes
\begin{align}
\overline{Y}_{\text{NR,off,im}}^{(z)}\left(d_{\max}\right)= & \frac{\sigma_{c}^{2}(2^{\bar{e}}-1)}{g(d_{\max})}\left(1-\frac{1}{1-e^{-\rho_{z}2d_{\max}}}\right)\nonumber \\
 & +\frac{1}{1-e^{-\rho_{z}2d_{\max}}}\int_{0}^{\sigma_{c}^{2}(2^{\bar{e}}-1)/g(d_{\max})}\left(-y\frac{\ln10}{10}\right)\left(-\frac{1}{y}\frac{10}{\ln10}\right)e^{-\rho_{z}2g_{\text{dB}}^{-1}\left(10\log_{10}\left(\frac{1}{y}\sigma_{c}^{2}\left(2^{\bar{e}}-1\right)\right)\right)}dy\nonumber \\
= & \frac{\sigma_{c}^{2}(2^{\bar{e}}-1)}{g(d_{\max})}\left(1-\frac{1}{1-e^{-\rho_{z}2d_{\max}}}\right)\nonumber \\
 & -\frac{1}{1-e^{-\rho_{z}2d_{\max}}}\frac{\ln10}{10}\int_{+\infty}^{g_{\text{dB}}\left(d_{\max}\right)}\frac{\sigma_{c}^{2}\left(2^{\bar{e}}-1\right)}{10^{y'/10}}e^{-\rho_{z}2g_{\text{dB}}^{-1}\left(y'\right)}dy'\nonumber \\
= & \frac{\sigma_{c}^{2}(2^{\bar{e}}-1)}{g(d_{\max})}\left(\frac{1}{1-e^{\rho_{z}2d_{\max}}}\right)+\frac{\sigma_{c}^{2}\left(2^{\bar{e}}-1\right)}{1-e^{-\rho_{z}2d_{\max}}}\frac{\ln10}{10}\int_{g_{\text{dB}}\left(d_{\max}\right)}^{+\infty}10^{-y'/10}e^{-\rho_{z}2g_{\text{dB}}^{-1}\left(y'\right)}dy',\label{eq:app4}
\end{align}
i.e., the desired Eq.~\eqref{eq:Y_OFF_IMM_mean}.

\subsection{Probability Distribution Function, Probability Density Function,
and average value of the transmit power used for delayed content delivery
through I2D}

The CDF of the random variable $Y_{\text{NR,non-off}}$ defined in
\eqref{eq:Y_NR_NON_OFF_DEFINITION} can be computed from the distribution
of the transmission distance from the eNodeB to the device, which
is uniformly distributed in the interval $[0,d_{\max}^{\text{(I2D)}}]$, or
$F_{D}\left(d\right)=\frac{1}{d_{\max}^{\text{(I2D)}}}d\cdot u_{[0,d_{\max}^{\text{(I2D)}}]}\left(d\right)$.
Specifically, we have:
\begin{align}
F_{Y,\text{NR,non-off}}\left(y\right) & =Pr\left(Y_{\text{NR,non-off}}\leq y\right)\nonumber\\
 & =Pr\left(\frac{1}{g(D)}\sigma_{c}^{2}\left(2^{\bar{e}}-1\right)\leq y\right)\nonumber\\
 & =Pr\left(g(D)\geq\frac{1}{y}\sigma_{c}^{2}\left(2^{\bar{e}}-1\right)\right)\nonumber\\
 & =Pr\left(g_{\text{dB}}(D)\geq10\log_{10}\left(\frac{1}{y}\sigma_{c}^{2}\left(2^{\bar{e}}-1\right)\right)\right)\nonumber\\
 & =Pr\left(D\leq g_{\text{dB}}^{-1}\left(10\log_{10}\left(\frac{1}{y}\sigma_{c}^{2}\left(2^{\bar{e}}-1\right)\right)\right)\right)
\end{align}
and, ultimately,
\begin{align}
F_{Y,\text{NR,non-off}}\left(y\right) & =\frac{1}{d_{\max}^{\text{(I2D)}}}g_{\text{dB}}^{-1}\left(10\log_{10}\left(\frac{1}{y}\sigma_{c}^{2}\left(2^{\bar{e}}-1\right)\right)\right)u_{[0,\sigma_{c}^{2}\left(2^{\bar{e}}-1\right)/g\left(d_{\max}^{\text{(I2D)}}\right)]}\left(y\right),
\end{align}
i.e., Eq.~\eqref{eq:Y_NON_off_CDF}.

For the sake of completeness, it is worth also computing the PDF of
$Y_{\text{NR,non-off}}$, even though it is not required to compute
its average value, since the integration by parts in the computation
of the average value gets rid of the PDF (see below). The PDF of $Y_{\text{NR,non-off}}$
is given by the first derivative of the CDF, i.e.:
\begin{align}
p_{Y,\text{NR,non-off}}(y) & =\frac{d}{dy}F_{Y,\text{non-off}}\left(y\middle|D\leq d_{\max}^{\text{(I2D)}}\right)\nonumber\\
 & =\frac{d}{dy}\left(\frac{1}{d_{\max}^{\text{(I2D)}}}g_{\text{dB}}^{-1}\left(10\log_{10}\left(\frac{1}{y}\sigma_{c}^{2}\left(2^{\bar{e}}-1\right)\right)\right)u_{[0,\sigma_{c}^{2}\left(2^{\bar{e}}-1\right)/g\left(d_{\max}^{\text{(I2D)}}\right)]}\left(y\right)\right)\nonumber\\
 & =\frac{1}{d_{\max}^{\text{(I2D)}}}\frac{d}{dy}\left(g_{\text{dB}}^{-1}\left(10\log_{10}\left(\frac{1}{y}\sigma_{c}^{2}\left(2^{\bar{e}}-1\right)\right)\right)\right)u_{[0,\sigma_{c}^{2}\left(2^{\bar{e}}-1\right)/g\left(d_{\max}^{\text{(I2D)}}\right)]}\left(y\right).
\end{align}
Applying the rule for the derivative of an inverse function $\frac{d}{dx}f^{-1}\left(x\right)=\frac{1}{f'\left(f^{-1}\left(x\right)\right)}$,
and the chain rule for the derivative of nested functions $\frac{d}{dx}f\left(g\left(x\right)\right)=f'\left(g\left(x\right)\right)g'\left(x\right)$,
we obtain
\begin{align}
p_{Y,\text{NR,non-off}}(y)= & \frac{1}{d_{\max}^{\text{(I2D)}}}\frac{1}{g_{\text{dB}}'\left(g_{\text{dB}}^{-1}\left(10\log_{10}\left(\frac{1}{y}\sigma_{c}^{2}\left(2^{\bar{e}}-1\right)\right)\right)\right)}\frac{d}{dy}\left(10\log_{10}\left(\frac{1}{y}\sigma_{c}^{2}\left(2^{\bar{e}}-1\right)\right)\right)\nonumber\\
 & \cdot u_{[0,\sigma_{c}^{2}\left(2^{\bar{e}}-1\right)/g\left(d_{\max}^{\text{(I2D)}}\right)]}\left(y\right)\nonumber\\
= & \frac{1}{d_{\max}^{\text{(I2D)}}}\frac{1}{g_{\text{dB}}'\left(g_{\text{dB}}^{-1}\left(10\log_{10}\left(\frac{1}{y}\sigma_{c}^{2}\left(2^{\bar{e}}-1\right)\right)\right)\right)}\frac{10}{\ln10}\frac{d}{dy}\left(\ln\left(\frac{1}{y}\sigma_{c}^{2}\left(2^{\bar{e}}-1\right)\right)\right)\nonumber\\
 & \cdot u_{[0,\sigma_{c}^{2}\left(2^{\bar{e}}-1\right)/g\left(d_{\max}^{\text{(I2D)}}\right)]}\left(y\right)\nonumber\\
= & \frac{1}{d_{\max}^{\text{(I2D)}}}\frac{1}{g_{\text{dB}}'\left(g_{\text{dB}}^{-1}\left(10\log_{10}\left(\frac{1}{y}\sigma_{c}^{2}\left(2^{\bar{e}}-1\right)\right)\right)\right)}\frac{10}{\ln10}\frac{y}{\sigma_{c}^{2}\left(2^{\bar{e}}-1\right)}\frac{d}{dy}\left(\frac{1}{y}\sigma_{c}^{2}\left(2^{\bar{e}}-1\right)\right)\nonumber\\
 & \cdot u_{[0,\sigma_{c}^{2}\left(2^{\bar{e}}-1\right)/g\left(d_{\max}^{\text{(I2D)}}\right)]}\left(y\right)\nonumber\\
= & -\frac{1}{y}\frac{1}{d_{\max}^{\text{(I2D)}}}\frac{1}{g_{\text{dB}}'\left(g_{\text{dB}}^{-1}\left(10\log_{10}\left(\frac{1}{y}\sigma_{c}^{2}\left(2^{\bar{e}}-1\right)\right)\right)\right)}\frac{10}{\ln10}u_{[0,\sigma_{c}^{2}\left(2^{\bar{e}}-1\right)/g\left(d_{\max}^{\text{(I2D)}}\right)]}\left(y\right)
\end{align}
Note that this expression is always a non-negative quantity, since
$g'\left(\cdot\right)$ is always negative, because the channel gain
is a decreasing function of its argument.

The average value of $Y_{\text{NR,non-off}}$ can be computed as follows
\begin{align}
\overline{Y}_{\text{NR,non-off}}\left(d_{\max}^{\text{(I2D)}}\right)= & \int_{-\infty}^{+\infty}y\cdot p_{Y,\text{NR,non-off}}\left(y\right)dy=\int_{0}^{\sigma_{c}^{2}\left(2^{\bar{e}}-1\right)/g\left(d_{\max}^{\text{(I2D)}}\right)}y\cdot p_{Y,\text{NR,non-off}}\left(y\right)dy\nonumber \\
= & \left[y\cdot F_{Y,\text{NR,non-off}}\left(y\right)\right]_{0}^{\frac{\sigma_{c}^{2}(2^{\bar{e}}-1)}{g(d_{\max}^{\text{(I2D)}})}}-\int_{0}^{\sigma_{c}^{2}(2^{\bar{e}}-1)/g(d_{\max}^{\text{(I2D)}})}\hspace{-20mm}F_{Y,\text{NR,non-off}}\left(y\right)dy\nonumber \\
= & \frac{\sigma_{c}^{2}(2^{\bar{e}}-1)}{g(d_{\max}^{\text{(I2D)}})}-\int_{0}^{\sigma_{c}^{2}(2^{\bar{e}}-1)/g(d_{\max}^{\text{(I2D)}})}\frac{1}{d_{\max}^{\text{(I2D)}}}g_{\text{dB}}^{-1}\left(10\log_{10}\left(\frac{1}{y}\sigma_{c}^{2}\left(2^{\bar{e}}-1\right)\right)\right)dy\nonumber \\
= & \frac{\sigma_{c}^{2}(2^{\bar{e}}-1)}{g(d_{\max}^{\text{(I2D)}})}-\frac{1}{d_{\max}^{\text{(I2D)}}}\int_{0}^{\sigma_{c}^{2}(2^{\bar{e}}-1)/g(d_{\max}^{\text{(I2D)}})}\hspace{-5mm}g_{\text{dB}}^{-1}\left(10\log_{10}\left(\frac{1}{y}\sigma_{c}^{2}\left(2^{\bar{e}}-1\right)\right)\right)dy.\label{eq:app}
\end{align}
~

We now apply the same change of the integration variable introduced in \eqref{eq:varchange} to compute \eqref{eq:app4}. In this way, \eqref{eq:app} becomes
\begin{align}
\overline{Y}_{\text{NR,non-off}}\left(d_{\max}^{\text{(I2D)}}\right)= & \frac{\sigma_{c}^{2}(2^{\bar{e}}-1)}{g(d_{\max}^{\text{(I2D)}})}-\frac{1}{d_{\max}^{\text{(I2D)}}}\int_{0}^{\frac{\sigma_{c}^{2}(2^{\bar{e}}-1)}{g(d_{\max}^{\text{(I2D)}})}}\left(-y\frac{\ln10}{10}\right)\left(-\frac{1}{y}\frac{10}{\ln10}\right)\nonumber\\
&\cdot g_{\text{dB}}^{-1}\left(10\log_{10}\left(\frac{1}{y}\sigma_{c}^{2}\left(2^{\bar{e}}-1\right)\right)\right)dy\nonumber\\
= & \frac{\sigma_{c}^{2}(2^{\bar{e}}-1)}{g(d_{\max}^{\text{(I2D)}})}+\frac{1}{d_{\max}^{\text{(I2D)}}}\frac{\ln10}{10}\int_{+\infty}^{g_{\text{dB}}\left(d_{\max}^{\text{(I2D)}}\right)}\frac{\sigma_{c}^{2}\left(2^{\bar{e}}-1\right)}{10^{y'/10}}g_{\text{dB}}^{-1}\left(y'\right)dy'\nonumber\\
= & \frac{\sigma_{c}^{2}(2^{\bar{e}}-1)}{g(d_{\max}^{\text{(I2D)}})}-\frac{\sigma_{c}^{2}\left(2^{\bar{e}}-1\right)}{d_{\max}^{\text{(I2D)}}}\frac{\ln10}{10}\int_{g_{\text{dB}}\left(d_{\max}^{\text{(I2D)}}\right)}^{+\infty}10^{-y'/10}g_{\text{dB}}^{-1}\left(y'\right)dy',
\end{align}
i.e., the desired result \eqref{eq:Y_NON_off_mean}.
\end{document}